\documentclass[aip,jcp,graphicx,12pt]{revtex4-1}

\usepackage{graphicx}
\usepackage{dcolumn}
\usepackage{bm}

\usepackage[utf8]{inputenc}
\usepackage[T1]{fontenc}
\usepackage{mathptmx}
\usepackage{etoolbox}  
\usepackage{hyperref}
\usepackage{url}            
\usepackage{booktabs}       
\usepackage{amsfonts}       
\usepackage{nicefrac}       
\usepackage{microtype}      
\usepackage{xcolor}         
\usepackage{multirow}
\usepackage{amsmath}
\usepackage{graphicx}
\usepackage{subcaption}
\usepackage{chemformula}
\usepackage{braket}
\usepackage{algorithmic}
\usepackage[ruled,vlined]{algorithm2e}
\usepackage{amsmath}
\usepackage{paralist}
\usepackage{xr}
\usepackage{cleveref} 
\makeatletter
\def\@email#1#2{%
 \endgroup
 \patchcmd{\titleblock@produce}
  {\frontmatter@RRAPformat}
  {\frontmatter@RRAPformat{\produce@RRAP{*#1\href{mailto:#2}{#2}}}\frontmatter@RRAPformat}
  {}{}
}%
\makeatother
\begin{document}

\preprint{AIP/JCP}

\title[FB-GNN-MBE]{Transferable FB-GNN-MBE Framework for 
Potential Energy Surfaces: 
Data-Adaptive Transfer Learning in Deep Learned Many-Body Expansion Theory} 
\author{Siqi Chen}
\affiliation{Department of Chemical and Biomolecular Engineering, University of Massachusetts Amherst, Amherst, MA 01003}
\author{Zhiqiang Wang}
\altaffiliation[Also at ]{Department of Electrical Engineering and Computer Science, Florida Atlantic University, Boca Raton, FL 33431}
\affiliation{Department of Chemistry, University of Massachusetts Amherst, Amherst, MA 01003}
\author{Yili Shen}
\altaffiliation[Also at ]{Computer Science and Engineering, University of Notre Dame, Notre Dame, IN 46556}
\affiliation{Manning College of Information and Computer Sciences, University of Massachusetts Amherst, Amherst, MA 01003}
\author{Xianqi Deng}
\altaffiliation[Also at ]{Department of Computer Science, University at Albany, State University of New York, Albany, NY 12222}
\affiliation{Department of Chemistry, University of Massachusetts Amherst, Amherst, MA 01003}
\author{Xi Cheng}
\affiliation{Department of Chemistry, University of Massachusetts Amherst, Amherst, MA 01003}
\author{Cheng-Wei Ju}
\altaffiliation[Also at ]{Division of Biology and Biological Engineering, California Institute of Technology, Pasadena, CA 91125}
\affiliation{Department of Chemistry, University of Massachusetts Amherst, Amherst, MA 01003}
\author{Jun Yi}
\altaffiliation[Also at ]{Department of Chemistry, Wake Forest University, Winston-Salem, NC 27109}
\affiliation{Department of Chemistry, University of Massachusetts Amherst, Amherst, MA 01003}
\author{Guo Ling}
\affiliation{Department of Chemistry, University of Massachusetts Amherst, Amherst, MA 01003}
\author{Dieaa Alhmoud}
\affiliation{Department of Chemistry, University of Massachusetts Amherst, Amherst, MA 01003}
\author{Hui Guan}
\affiliation{Manning College of Information and Computer Sciences, University of Massachusetts Amherst, Amherst, MA 01003}
\author{Zhou Lin}
\email{zhoulin@umass.edu}
\homepage{https://websites.umass.edu/zlinqcgroup/}
\affiliation{Department of Chemistry, University of Massachusetts Amherst, Amherst, MA 01003}

\date{\today}
\maketitle
\clearpage
\section*{abstract}
Mechanistic understanding and rational design of complex chemical systems depend on fast and accurate predictions of electronic structures beyond individual building blocks. 
However, if the system exceeds hundreds of atoms, first-principles quantum mechanical (QM) modeling becomes impractical. 
In this study, we developed FB-GNN-MBE by integrating a fragment-based graph neural network (FB-GNN) into the many-body expansion (MBE) theory and demonstrated its capacity 
to reproduce first-principles potential energy surfaces (PES) for hierarchically structured systems with manageable accuracy, complexity, and interpretability. 
Specifically, we divided the entire system into basic building blocks (fragments), evaluated their one-fragment energies using a 
QM model, and addressed many-fragment interactions using the structure–property relationships trained by FB-GNNs.
Our investigation shows that FB-GNN-MBE achieves chemical accuracy in predicting two-fragment (2B) and three-fragment (3B) energies across water, phenol, and mixture benchmarks, as well as the one-dimensional dissociation curves of water and phenol dimers.
To transfer the success of FB-GNN-MBE across various systems with minimal computational costs and data demands, we developed and validated a teacher--student learning protocol. 
A heavy-weight FB-GNN trained on a mixed-density water cluster ensemble (teacher) distills its learned knowledge and passes it to a light-weight GNN (student), which is later fine-tuned on a normal-density $(\ch{H2O})_{21}$ cluster ensemble.
This transfer learning strategy resulted in efficient and accurate prediction of 2B and 3B energies for variously sized water clusters without retraining.
Our transferable FB-GNN-MBE framework outperformed 
conventional non-FB-GNN-based models and 
\textcolor{black}{provided a scalable and accurate route toward interaction energies of large molecular assemblies}.

\clearpage

\section{Introduction}
\label{sec:introduction}

Non-covalent intermolecular interactions play an essential but complex role in chemical phenomena 
in the condensed phase.
\cite{cisneros2016modeling,doi:10.1021/acs.chemrev.5b00584,https://doi.org/10.1002/anie.202316364} 
For example, the dynamic hydrogen bond network among water and other chemical motifs governs various behaviors, such as protein folding,\cite{levy2006water,sumi2021water,doi:10.1073/pnas.1411798111} surface electroreduction,\cite{doi:10.1021/acscatal.3c01223,doi:10.1021/acscatal.3c05880,tian2025electrochemical} and lattice vibration.\cite{desiraju2002hydrogen,doi:10.1021/acsomega.4c05344,D4EE01219D}
Accurately modeling non-covalent interactions is essential for understanding and predicting aggregate quantum mechanical (QM) 
properties that modulate these phenomena, and it requires a fast and reliable way to evaluate potential energy surfaces (PESs) that capture 
intermolecular and intramolecular interactions and structural variations. 
A typical first-principles QM method, such as coupled-cluster singles, doubles, and perturbative triples [CCSD(T)],\cite{RAGHAVACHARI1989479} second-order M{\o}ller--Plesset perturbation theory (MP2),
\cite{moller1934note} and even density functional theory (DFT),\cite{PhysRev.136.B864,PhysRev.140.A1133} is popular for constructing the PES, but it suffers from the high 
computational complexity when applied 
to large-scale \textit{ab initio} molecular dynamics (MD) simulations.\cite{doi:https://doi.org/10.1002/9781119019572,RevModPhys.71.1085}
A classical force field with fitted parameters, on the other hand,
offers high speed in MD at the expense of chemical fidelity, such as the missing charge fluctuations in dynamic hydrogen bond networks.\cite{WARSHEL1976227,xiong2022fast,rick2023effects,heindel2023many}

To speed up high-fidelity PES calculations for a complex system such as a large water or phenol cluster, fragment-based ``divide-and-conquer'' methods have been developed.\cite{doi:10.1021/acs.accounts.6b00356,10.1063/1.5126216}
Among them, the many-body expansion (MBE) theory decomposes the system into fragments and expands its total energy (or another aggregate property) into one-body (1B, one-fragment) terms from isolated fragments and many-body ($n$B, many-fragment) corrections that decay rapidly in magnitude.\cite{doi:10.1021/ct700223r,doi:10.1021/ar500119q,heindel2022many}
While reducing computational complexity, MBE provides clear physical insight into many-fragment interactions in electronic structures.
A few groups have made major advances in the MBE methodology, covering both the static and dynamic behaviors of condensed phase systems. 
For example, Herbert and coworkers introduced generalized MBE (GMBE) to handle overlapping fragments and energy‑screened MBE to consider only important many‑body terms; \cite{10.1063/1.4742816,doi:10.1021/ct300985h,doi:10.1021/jz401368u,10.1063/1.4885846,10.1063/1.4947087,doi:10.1021/acs.jctc.5b00955,10.1063/1.4986110,doi:10.1021/acs.jctc.9b01095,10.1063/5.0174293}
Xantheas and coworkers leveraged MBE to build quantitative 
PESs and embedded them in MD simulations to capture electronic and nuclear quantum effects.\cite{heindel2020many,doi:10.1021/acs.jctc.0c01309,doi:10.1021/acs.jctc.1c00780,10.1063/5.0095335,D1CP00409C,10.1063/5.0095739,doi:10.1021/acs.jctc.3c00575,herman2023extensive,doi:10.1021/acs.jpclett.2c03822,10.1063/5.0094598,PhysRevC.107.044004}
However, a QM-based MBE approach remains computationally costly for extended systems because many-body terms still 
proliferate.

The rapid evolution of artificial intelligence for science provides a new, ever-growing toolbox to accelerate QM approaches in general\cite{brockherde2017bypassing,ryabov2020neural,ramakrishnan2015big} and in PES calculations.\cite{behler2007generalized,PhysRevLett.104.136403,PhysRevLett.120.143001}
For example, Parkhill and coworkers implemented neural networks (NNs) in MBE (NN-MBE) and expanded the total energy of a methanol (\ch{CH3OH}) cluster to up to the 3B terms.
Compared to MP2 calculations, NN-MBE delivers mean absolute errors (MAEs) of 9.79 and 12.55 cal/mol for 2B and 3B energies, respectively.\cite{yao2017many} 
However, because molecular features and parameters are not uniquely mapped to physical quantities, traditional NNs lack physical interpretability and transfer weakly among systems.
\cite{10.1145/3233231,rudin2019stop,9007737}
Graph neural networks (GNNs) with the message passing schemes,\cite{kipf2016semi,NIPS2017_5dd9db5e} on the other hand, allows their nodes and edges to align naturally with atoms and bonds, and can therefore encode geometric and directional information with chemical significance and support end-to-end learning of molecular properties under permutation, translation, and rotation equivariance.\cite{kearnes2016molecular,10.5555/2969442.2969488,chen2019graph}
Representative GNN models, which exhibit great predictive powers in molecular properties, include SchNet\cite{NIPS2017_303ed4c6,schutt2018schnet} and PhysNet\cite{unke2019physnet} with continuous filters, DimeNet\cite{gasteiger2020directional} and DimeNet++\cite{gasteiger2020fast} with directional message passing, EGNN and SEGNN \cite{satorras2021n, brandstetter2021geometric} with $E(3)$‑equivariance, ViSNet \cite{wang2024enhancing} with directional vector--scalar interactions, and MACE\cite{batatia2022mace} with 
equivariant interatomic potentials.
However, most GNN models treat all atoms uniformly and ignore chemical hierarchy, which limits their performance on complex systems built from recurring fragments and motivates the development of fragment-based GNNs (FB-GNNs) to account for many‑fragment properties. 
Outstanding examples of FB-GNNs include MXMNet with 
two-layer multiplex graph representations inspired by molecular mechanics,\cite{zhang2020molecular} PAMNet with physics-informed bias to model local and non-local interactions differently,\cite{zhang2023universal} SubGNN with explicit subgraph topology features,\cite{alsentzer2020subgraph} and FragGraph with fragment-based fingerprints embedded in molecular graphs.\cite{doi:10.1021/acs.jpca.1c06152} 
All of them represent fragments as interacting subgraphs or local graphs and couple local (intrafragment) and global (interfragment) message passing.\cite{gilmer2017neural} 
With a heterogeneous structure, such a hierarchic graph captures short‑range bonding interactions, longer‑range non-covalent interactions, and high-order many‑fragemnts interactions, and constructs generalizable models across sizes and motifs.\cite{jiang2021contrastive,wang2022survey}

To improve the scalability and efficiency of MBE and preserve its 
additivity and interpretability, we integrated FB-GNNs into the original MBE framework (FB-GNN-MBE).\cite{chen2024integrating}
Specifically, we implemented MXMNet\cite{zhang2020molecular} and PAMNet\cite{zhang2023universal} as our backbone FB-GNN models, because their multiplex global--local message passing schemes are appropriate for learning both short-range intrafragment interactions and long-range interfragment ones.
In the first version of this hybrid approach, we expanded the total ground state energy of a complex system up to 3B terms, evaluating the inexpensive 1B energies using MP2 or DFT, and leveraging FB-GNNs to learn the complex, configuration-dependent 2B and 3B corrections within the chemical accuracy. 
In addition, we addressed the lack of flexibility and transferability of a machine-learned PES when extrapolated to under-sampled regions where system configurations deviate from those in the training set, without expanding the training set through high-level QM calculations.\cite{behler2007generalized,bowman2022delta,muniz2021vapor,10.1063/1.473987,10.1063/1.473864,collins2002molecular}
To address these issues while maintaining a low computational cost, we introduced a teacher--student knowledge distillation protocol, \cite{stanton2021does,liu2023network} in which chemical and physical principles in non-covalent interactions are distilled as compact softened outputs or learned representations from a large existing dataset using a heavy-weight, pre-trained ``teacher'' model and passed down to a light-weight, fine-tuned ``student'' model based on a minimal add-on dataset. 
The distillation step is in particular beneficial for small clusters because it can help avoid catastrophic overfitting and forgetting.
\cite{pratt1992discriminability,yosinski2014transferable,smith2019approaching,Hu2020Strategies,hinton2015distilling,wu2022knowledge}  
In the present study, we applied PAMNet\cite{zhang2023universal} as the teacher model and four popular GNNs (DimeNet,\cite{gasteiger2020directional} DimeNet++,\cite{gasteiger2020fast} ViSNet,\cite{wang2024enhancing} and SchNet\cite{NIPS2017_303ed4c6,schutt2018schnet}) as the student models.

The remainder of the paper is organized as follows. 
Section \ref{sec:method} describes the theoretical foundations and computational details of the FB-GNN-MBE framework and the teacher–student protocol, along with their datasets. 
Section \ref{sec:results} presents the results to assess the performance of our models in water and phenol clusters.
Section \ref{sec:conclusion} summarizes our findings and discusses future directions.

\section{Methods}
\label{sec:method}

\subsection{Many-Body Expansions}
\label{sec:fb-gnn-mbe}

\begin{figure}[!h]
\centering
\includegraphics[width=0.5\textwidth]{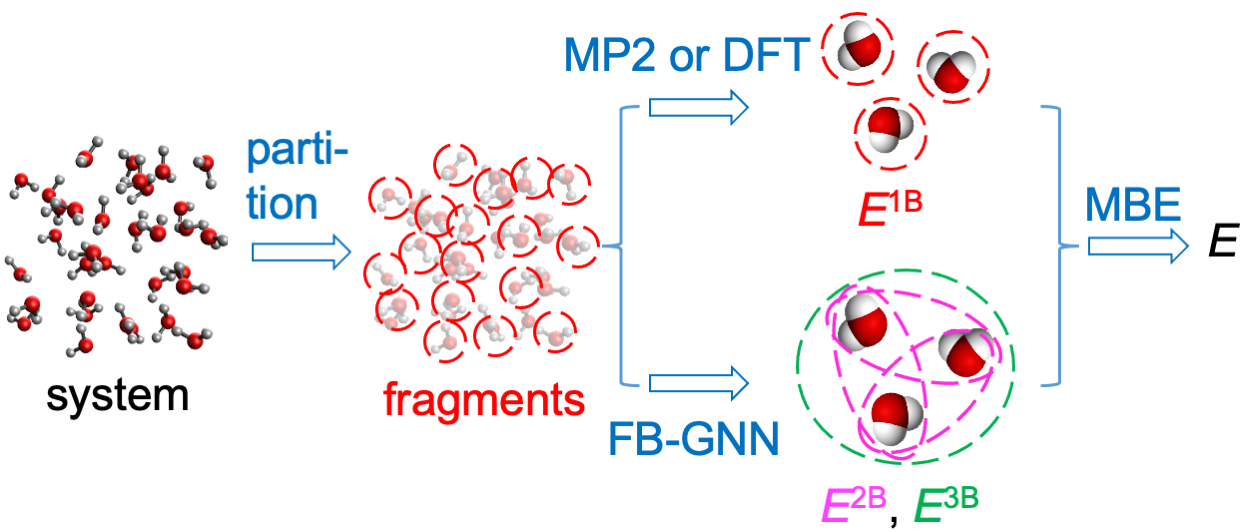}
    \caption{Schematic strategy of our FB-GNN-MBE framework, using a water cluster as an illustrative example, modified from our earlier study.\cite{chen2024integrating} 1B energies are calculated using MP2 or DFT, and 2B and 3B corrections are obtained from FB-GNN-trained structure--property relationships.}
    \label{fig:1-mbe}
\end{figure}

FB-GNN-MBE is established on top of MBE, which decomposes the total ground state energy of an $N$-fragment system ($E$) into 1B, 2B, 3B, and $\ldots$ terms.\cite{heindel2022many} 
Due to the smaller magnitudes of the non-covalent interactions and the need to control the computational complexity, the expansion is truncated at the 3B level in the present study.
As such
\begin{equation}
E \simeq \sum_{i=1}^{N} E_{i}^{\text{1B}} + \sum_{1\le i<j\le N}^{N} E_{ij}^{\text{2B}} + \sum_{1\le i<j<k\le N}^{N} E_{ijk}^{\text{3B}} 
\label{eq:mbe}
\end{equation}
where each 1B term ($E_{i}^{\text{1B}}$) is the energy of the $i^\text{th}$ isolated fragment (monomer), and the 2B and 3B corrections ($E_{ij}^{\text{2B}}$ and $E_{ijk}^{\text{3B}}$) are defined recursively as the energy difference between the full dimer or trimer energy and the sum of its constituent lower-order terms. 
As such 
\begin{align}
E^\text{1B}_{i} & = E_{i} \label{eq:1b} \\
E^\text{2B}_{ij} & = E_{ij} - E^\text{1B}_{i} - E^\text{1B}_{j} \label{eq:2b}\\
E^\text{3B}_{ijk} & = E_{ijk} - E^\text{2B}_{ij} - E^\text{2B}_{ik} - E^\text{2B}_{jk} - E^\text{1B}_{i} - E^\text{1B}_{j} - E^\text{1B}_{k} \label{eq:3b} 
\end{align}
Our hybrid FB-GNN-MBE scheme uses MP2 or DFT methods for the computationally inexpensive 1B terms, and employs FB-GNNs to learn and predict the more demanding 2B and 3B corrections directly from the instantaneous geometric configurations of the dimers and trimers (Figure \ref{fig:1-mbe}).

\subsection{Fragment-Based Graph Neural Networks}
\begin{figure}[!ht]
    \centering
 \includegraphics[width=\textwidth]{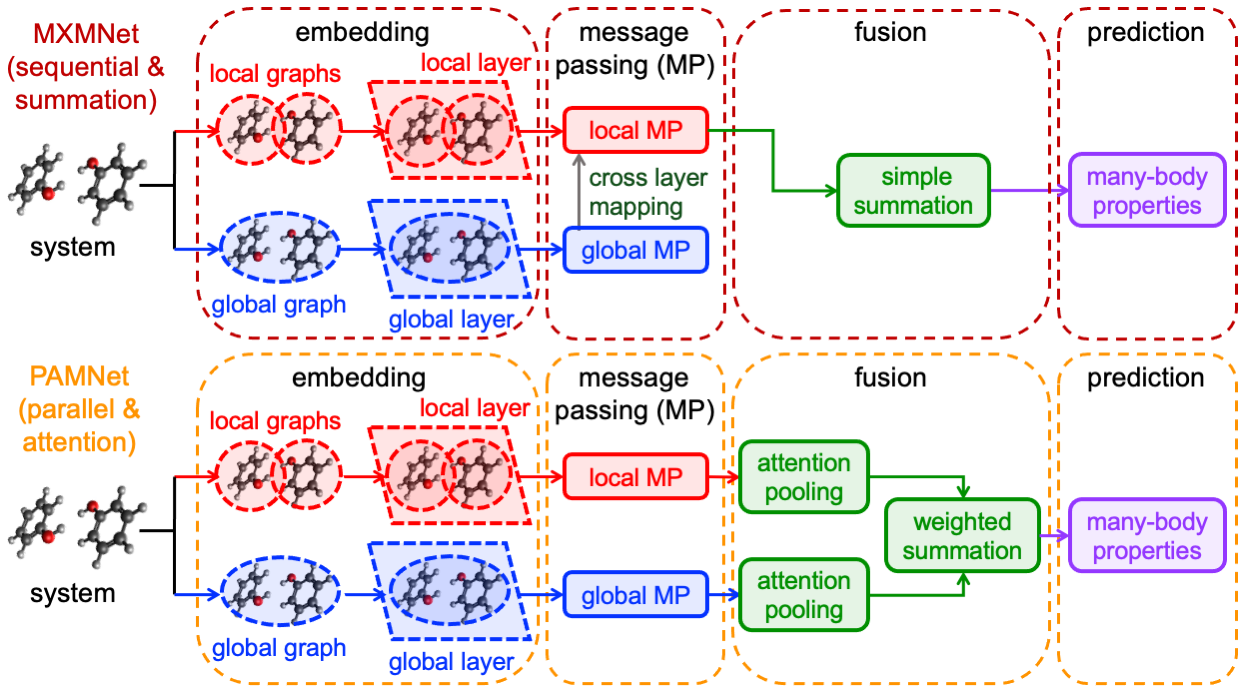}
    \caption{Schematic designs of MXMNet (top) and PAMNet (bottom) to model a hierarchic chemical system, created based on the descriptions of Xie and coworkers\cite{zhang2020molecular,zhang2023universal} and modified from our earlier study.\cite{chen2024integrating}}
    \label{fig:2-fbgnn} 
\end{figure}

The backbone methods of our FB-GNNs, MXMNet\cite{zhang2020molecular} and PAMNet,\cite{zhang2023universal} effectively capture both short- 
and long-range 
interactions through their multiplex global--local architecture [$\mathcal{G} = ({\mathcal{G}_\text{g},\mathcal{G}_\text{l}})$], with salient architecture information 
as described here. 
Both models represent the entire system as a global graph ($\mathcal{G}_{g}$) and the network of pre-defined building blocks (or fragments) as local graphs ($\mathcal{G}_{l}$).
They also represent short-range interatomic interactions as local edges within a local graph ($\mathcal{E}_{l}$) , and long-range interfragment interactions as global edges between local graphs ($\mathcal{E}_{g}$). 
Cross layer mapping and message passing algorithms\cite{gilmer2017neural} integrate all information from the global and local layers (Figure \ref{fig:2-fbgnn}). 

MXMNet and PAMNet each comprise five modules. 
They share identical embedding and prediction modules, but differ 
in how they handle 
message passing, attention pooling, and feature fusion. 
The embedding module utilizes atom-centered descriptors to characterize the local chemical environment. 
In particular, they apply radial basis functions (RBFs)\cite{schutt2018schnet,gasteiger2020directional}
to encode interatomic distances and spherical basis functions (SBFs)\cite{liu2022spherical,gasteiger2021gemnet,gasteiger2020directional} to capture angular information within atomic triplets.
This physics-informed representation scheme ensures $E(3)$-invariance\cite{batzner20223,batatia2022mace} and structural 
awareness, allowing us to directly implement the system configuration within the global and local graph representations for 
learning complex molecular patterns using GNN layers.
Following the embedding module, both models present a message passing module and a fusion module to integrate global and local information and preserve hierarchical, multi-scale features. 
MXMNet employs a sequential scheme, that starts with global message passing, followed by local message passing and cross layer mapping, and concludes with the fusion of global and local information via simple accumulation. 
The cross layer mapping interactively updates node embeddings and local graph representations.
PAMNet executes global and local message passing in parallel and fuses their outputs using an attention pooling mechanism {rather than using a cross layer mapping}. This treatment not only enhances the model's capacity to capture key molecular features, but also simplifies the process.
{In the end, the prediction module leverages the final node embeddings to predict fragment-specific and system-wide properties, especially 2B and 3B energies in the present study.}

In the present study, the robustness of both MXMNet and PAMNet was improved by implementing numerical clamping into MXMNet, updating the linear algebra libraries in PAMNet, and equipping both models with charge configurations for future studies.
Both models were trained using the $\mathcal{L}_1$ loss function, which is equivalent to the mean absolute error (MAE) of the predicted 2B or 3B energies and was chosen because it directly corresponds to the evaluation metrics and yields errors in a physically interpretable energy unit,
and avoids abrupt fluctuations in the gradient when occasional large errors appear. As such
\begin{equation}
\mathcal{L}_1 = \frac{1}{N}\sum_{i=1} \lvert E_{\text{predict},i} - E_{\text{true},i} \rvert 
\end{equation}
{in which $E_{\text{predict},i}$ and $E_{\text{true},i}$ represent the predicted and true values of a 2B or 3B energy, and $N$ represents the number of data points in the training set.}
We provide the pseudo-algorithms in {the Supplementary Materials (SM).} 

\subsection{Multi-Stage Training Strategy for Mixed-Density Datasets}
\label{sec:staged-training}
\begin{figure}[!ht]
    \centering
    \includegraphics[width=0.5\textwidth]{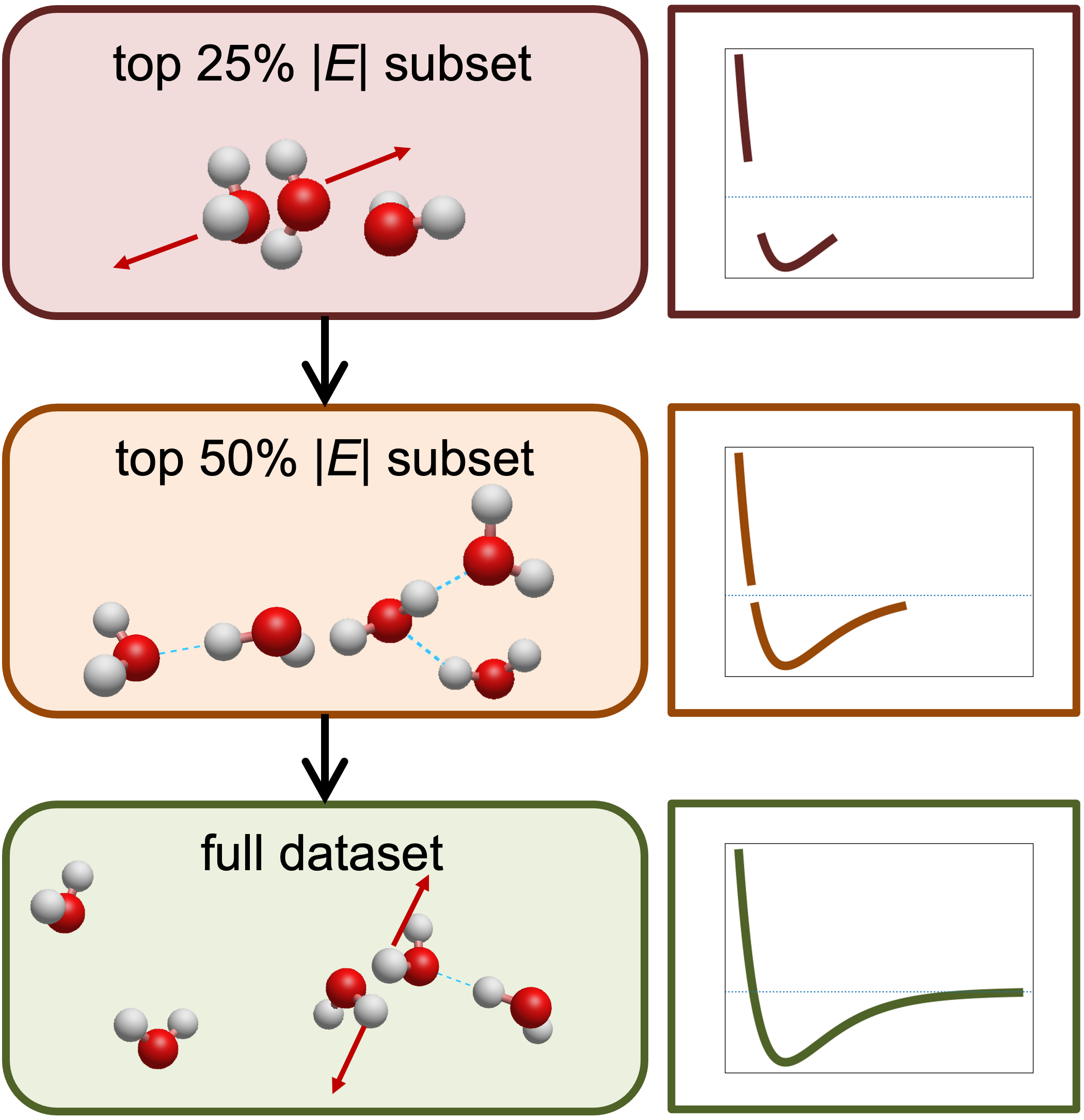}
    \caption{Multi-stage training strategy for low- and mixed-density datasets 
    by progressively learning from the high-energy region to the full-energy landscape.}
    \label{fig:3-training} 
\end{figure}

Due to a large number of near-zero 2B and 3B energies in the low- and mixed-density datasets (class imbalance), FB-GNN-MBE may bias toward excessive predictions of zero energies 
and fail to capture structure-dependent relationships.
To address this issue, a multi-stage training strategy based on curriculum learning\cite{bengio2009curriculum} was introduced, so that training data are used in segments, from simpler to more complex subsets, mimicking human and animal learning processes. 
In the present study, the dataset was reorganized based on the magnitudes of the target 2B or 3B energies.
The model was first pre-trained on a high-energy subset with top 25\% magnitudes of 2B or 3B energies to capture the most repulsive and most attractive regions 
of the PES, then trained on a medium- and high-energy subset with top 50\% magnitudes to balance with weaker repulsions and attractions, 
and finally fine-tuned with the full dataset, including configurations with all near-zero 2B or 3B energies (Figure \ref{fig:3-training}).

\subsection{Teacher--Student Knowledge Distillation Protocol on 
Under-Sampled Regions}
\label{sec:teacher-student}

\subsubsection{Teacher and Student Models}
\begin{figure}[!ht]
    \centering
 \includegraphics[width=0.5\textwidth]{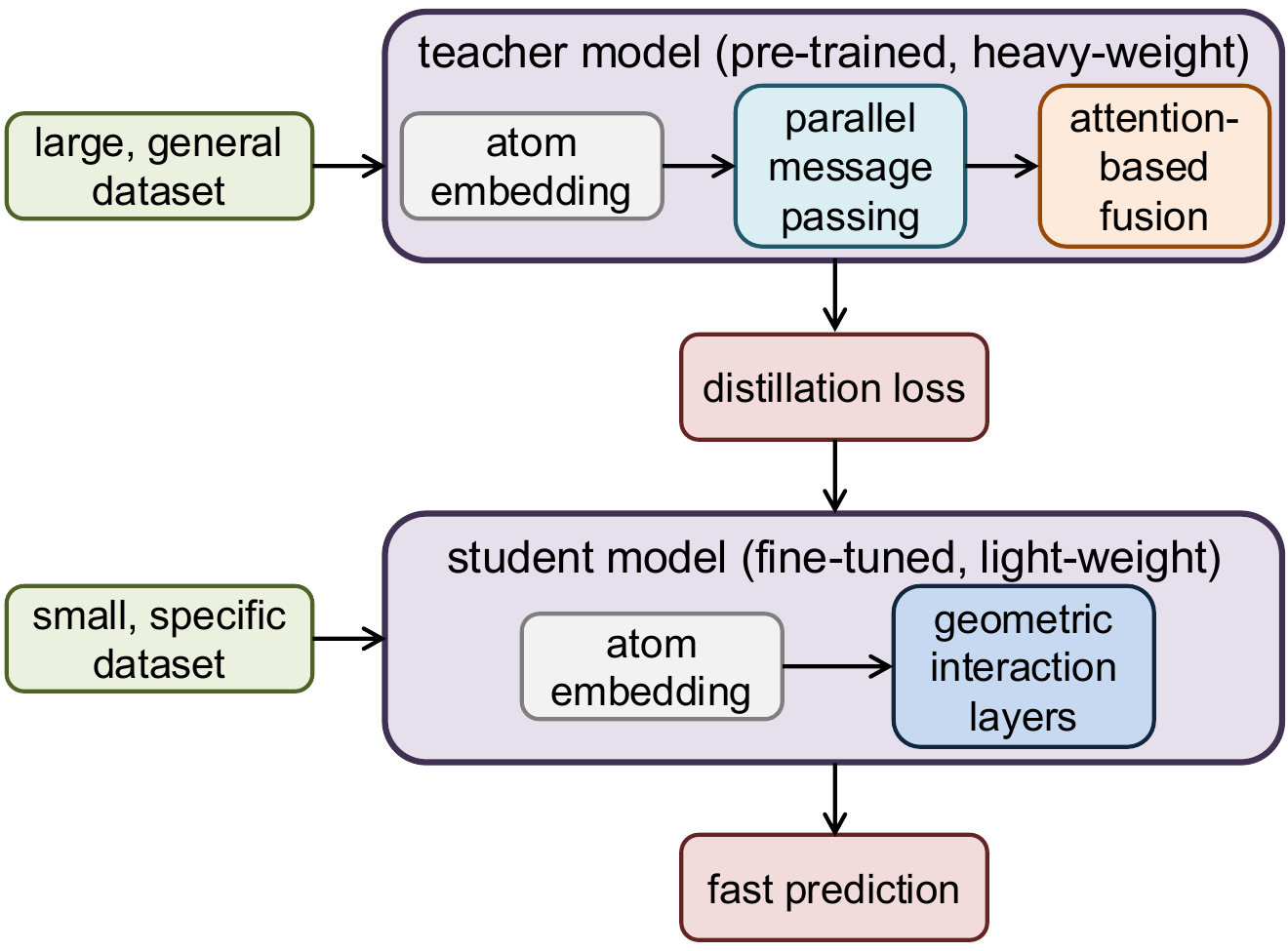}
    \caption{Teacher--student knowledge distillation protocol for under-sampled configurations 
    in which a pre-trained heavy-weight teacher model supervises a to-be-fine-tuned light-weight student model via knowledge distillation.}
    \label{fig:4-teacher-student} 
\end{figure}

To improve the transferability of the FB-GNN-MBE model to unseen configurations in under-sampled regions and promote 
its practical utility over a large configurational space (various cluster sizes, densities, and geometries), a teacher--student knowledge distillation protocol was employed.\cite{hinton2015distilling,wu2022knowledge}
The teacher--student protocol pre-trains modified PAMNet\cite{zhang2023universal} as a high-capacity teacher model using a large dataset of water clusters (explained in Section \ref{sec:system-dataset})\cite{gor2011matrix,zhang2016molecular} and extracts the general structure--property relationships from the training results.
The teacher serves as a repository of the structure-dependent patterns of intermolecular interactions, in particular, hydrogen bonds and van der Waals interactions, which are used to generate targets for distillation.
The protocol then passes this knowledge to a conventional non-fragment-based GNN (non-FB-GNN) as a light-weight student model 
and fine-tunes it with a small dataset to make sure it is more robust over noises (Figure \ref{fig:4-teacher-student}). 
Students were selected to span a range of representational complexity and geometric expressivity to examine how the richness of the features helps them learn and apply the distilled knowledge.
They include SchNet\cite{NIPS2017_303ed4c6,schutt2018schnet} which models pairwise distance-based interactions through pure distance encoding (RBF), DimeNet\cite{gasteiger2020directional} and DimeNet++,\cite{gasteiger2020fast} which model both distance- and angle-dependent interactions through distance and spherical encodings (RBF + SBF), and ViSNet\cite{wang2024enhancing} which further incorporates the rotation-equivariant (vector) representations to capture 3D geometry better.

\subsubsection{Knowledge Distillation} 
In the knowledge distillation step, the student model is trained to imitate the teacher model across many geometric configurations, using the teacher's outputs (predictions) 
as soft distillation targets. 
The distillation loss function ($\mathcal{L}_{\text{distill}}$) is a weighted sum of an energy‑matching term ($\mathcal{L}_{\text{energy}}$, as the mean squared error (MSE) in the predicted energies between the student and the teacher) and an optional feature‑matching term ($\mathcal{L}_{\text{feature}}$, as the MSE in the corresponding feature representations). As such,
\begin{align}
\mathcal{L}_{\text{distill}} & = \mathcal{L}_{\text{energy}} + \lambda \mathcal{L}_{\text{feature}} \\
\mathcal{L}_{\text{energy}} & = \frac{1}{N}\sum_{i=1}^N (E_{\text{student},i} - E_{\text{teacher},i})^2\\
\mathcal{L}_{\text{feature}} & = \frac{1}{N}\sum_{i=1}^N \lvert\lvert \mathbf{h}_{\text{student},i} - \mathbf{h}_{\text{teacher},i}\rvert\rvert^2
\end{align}
$\mathcal{L}_{\text{feature}}$ measures the discrepancy in graph embeddings ($\mathbf{h}$) and encourages the student to improve its understanding of the teacher's method, leading to better generalization, 
but is included only with a small value of $\lambda$ as a soft guide when the teacher and the student share comparable features.\cite{romero2014fitnets,hinton2015distilling} 

\subsubsection{Fine-Tuning} 
In the follow-up fine-tuning step, the student model is calibrated on a smaller dataset from the target domain to remove bias inherited from the teacher model due to the difference in data distributions. 
The fine-tuning loss function ($\mathcal{L}_{\text{fine-tune}}$) is the MAE of the normalized 2B or 3B energies (${\tilde{E}}$). 
As such, 
\begin{align}
\mathcal{L}_{\text{fine-tune}} & = \frac{1}{N} \sum_{i=1}^N 
\lvert {\tilde{E}}_{\text{student},i}
- {\tilde{E}}_{\text{true},i} \lvert 
\\
\tilde{E}_{\text{true},i} & = \frac{E_{\text{true},i} - E_{\min}}{E_{\max} - E_{\min}}
\end{align}
where $E_{\max}$ and $E_{\min}$ are the maximum and minimum energies taken from the training set. 
The original energy scale will be recovered when we evaluate the model performance. 
Fine-tuning is carried out with a reduced learning rate and an adaptive scheduler to adjust the model parameters for the high-fidelity target dataset without destroying the pre-trained features learned from the teacher model.\cite{li2017learning,ramakrishnan2015big}

\subsection{System and Dataset Generation}
\label{sec:system-dataset}

\begin{table}[h!]
\centering
\caption{Parameters and Details of Each Dataset.} 
\label{tab:md}
\begin{tabular}{c|c|c|c|c|c|c} 
\hline\hline
dataset & $\rho/\rho_\text{true}$ & $T$ (K) & monomers & dimers & trimers & training:validation:test \\ 
\hline
\midrule
\multicolumn{7}{l}{\textcolor{black}{(a) Teacher model: large-scale (pre-)training}} \\
\midrule
$\ch{(H2O)}_{17}$ & $0.5\times$ & 370 & 6,256 & 50,048 & 100,000 & 80:5:15\\ 
$\ch{(H2O)}_{33}$ & $1.0\times$ & 370 & 3,300 & 52,800 & 100,000 & 80:5:15\\ 
$\ch{(H2O)}_{50}$ & $1.5\times$ & 370 & 2,000 & 49,000 & 100,000 & 80:5:15\\ 
$\ch{(H2O)}_{67}$ & $2.0\times$ & 370 & 1,474 & 48,643 & 100,000 & 80:5:15\\ 
$\ch{(C6H5OH)}_{10}$ & $2.0\times$ & 360 & 10,010 & 45,045 & 120,120 & 80:5:15\\ 
$\ch{(H2O)_{10}:(C6H5OH)_{10}}$ & $2.0\times$ & 694 & 40,020  & 190,000 & 228,000 & 80:5:15\\ 
\midrule
\multicolumn{7}{l}{\textcolor{black}{(b) Student models: transfer learning [(\ch{H2O})$_{21}$] and external testing (others)}} \\
\midrule
$\ch{(H2O)_{7}}$ & $1.0\times$ & - & 7 & 21 & 35 & 0:0:100\\ 
$\ch{(H2O)_{10}}$ & $1.0\times$ & - & 10 & 45 & 120 & 0:0:100\\ 
$\ch{(H2O)_{13}}$ & $1.0\times$ & - & 13 & 78 & 286 & 0:0:100\\ 
$\ch{(H2O)_{16}}$ & $1.0\times$ & - & 16 & 120 & 560 & 0:0:100\\ 
$\ch{(H2O)_{21}}$ & $1.0\times$ & - & 21 & 210 & 1,330 & 90:10:0\\ 
\midrule
\multicolumn{7}{l}{\textcolor{black}{(c) 1D PES benchmark}} \\
\midrule
$\ch{(H2O)_{67}}$ & $2.0\times$ & 370 & 67 & 2,211 & 47,905 & 50:10:40 
\\
\hline\hline
\end{tabular}
\end{table}

\subsubsection{Summary of Datasets}
To provide a proof-of-concept for our FB-GNN-MBE framework and assess its robustness, accuracy, and efficiency in reproducing first-principles full-dimensional (FD) and one-dimensional (1D) PESs,
multiple benchmark systems were established using three types of molecular clusters with synergistic effects from hydrogen bonds and van der Waals interactions, including pure water clusters [(\ch{H2O})$_n$], pure phenol clusters [(\ch{C6H5OH})$_n$], and 
1:1 water:phenol mixture clusters [(\ch{H2O})$_n$:(\ch{C6H5OH})$_n$] (Table \ref{tab:md}). 
For each system, every single water or phenol molecule is treated as a fragment. 
These datasets were utilized for different purposes.

\subsubsection{Large Datasets for Teacher Model}
To assess the overall performance of FB-GNN-MBE on the 2B and 3B energies with diverse hydrogen bonds and van der Waals motifs and establish the teacher model, classical periodic molecular dynamics (MD) simulations were performed in the canonical ensemble (constant $NVT$) using GROMACS\cite{abraham2015gromacs} to sample the possible configurations of (\ch{H2O})$_{17}$, (\ch{H2O})$_{33}$, (\ch{H2O})$_{50}$, (\ch{H2O})$_{67}$, 
(\ch{C6H5OH})$_{10}$, and (\ch{H2O})$_{10}$:(\ch{C6H5OH})$_{10}$ at the desired densities ($\rho$), with initial configurations from PACKMOL.\cite{martinez2009packmol} 
For every snapshot, relevant monomers, dimers, and trimers, which span a wide range of geometries and non-covalent interactions, were extracted and calculated for the corresponding dataset. 
Each dataset was split randomly into 80:5:15 ratio for training, validation, and test sets, respectively.

\subsubsection{Small Datasets for Student Models}
To establish and assess the fine-tuned student models, a large (\ch{H2O})$_{21}$ cluster, the smallest water droplet showing bulk behaviors,\cite{D0SC05785A} was employed as the fine-tuning set, which was split randomly into 90:10 ratio for training and validation sets, respectively.
Smaller (\ch{H2O})$_{7}$, (\ch{H2O})$_{10}$, (\ch{H2O})$_{13}$, and (\ch{H2O})$_{16}$ clusters were utilized as the test sets. 

\subsubsection{Small Dataset for One-Dimensional Potential Energy Surface}
To evaluate the model capacity in reproducing the 1D dissociation curve over the $\ch{O-O}$ distances, 
one additional fine-tuning set was extracted from an unused snapshot in double-density MD simulations of $\ch{(H2O)}_{67}$ to make sure that it captures the repulsive walls (short \ch{O-O} distances), the equilibrium wells (medium \ch{O-O} distances), and the dissociation tails (long \ch{O-O} distances).\cite{rocken2025enhancing,radova2025fine,contant2024assessing,folmsbee2021evaluation,andreichev2025design}. 
This dataset was randomly split into 50:10:40 ratio for training, validation, and test sets, respectively.

\subsubsection{Computational Details}
Water clusters were calculated using MP2  
along with the aug-cc-pVDZ basis set,\cite{moller1934note,dunning1989gaussian} and phenol clusters and water:phenol clusters were calculated using DFT along with the $\omega$B97X-D3 exchange--correlation (XC) functional \cite{chai2008long,grimme2010consistent} and the 6-311+G(d,p) basis set.\cite{krishnan1980self} 
All calculations were performed using the Q-Chem 6.2  package.\cite{epifanovsky2021software}
Both MP2 and DFT have been confirmed 
to capture effective long-range interactions in hydrogen bonds and van der Waals interactions (especially $\pi$--$\pi$ stacking).\cite{jurevcka2006benchmark,goerigk2011thorough} 
\section{Results and Discussion}
\label{sec:results}

\subsection{Overall Performance of FB-GNN-MBE} 
\label{sec:overall-performance}

\begin{figure}[!h]
\centering
\includegraphics[width=0.5\textwidth]{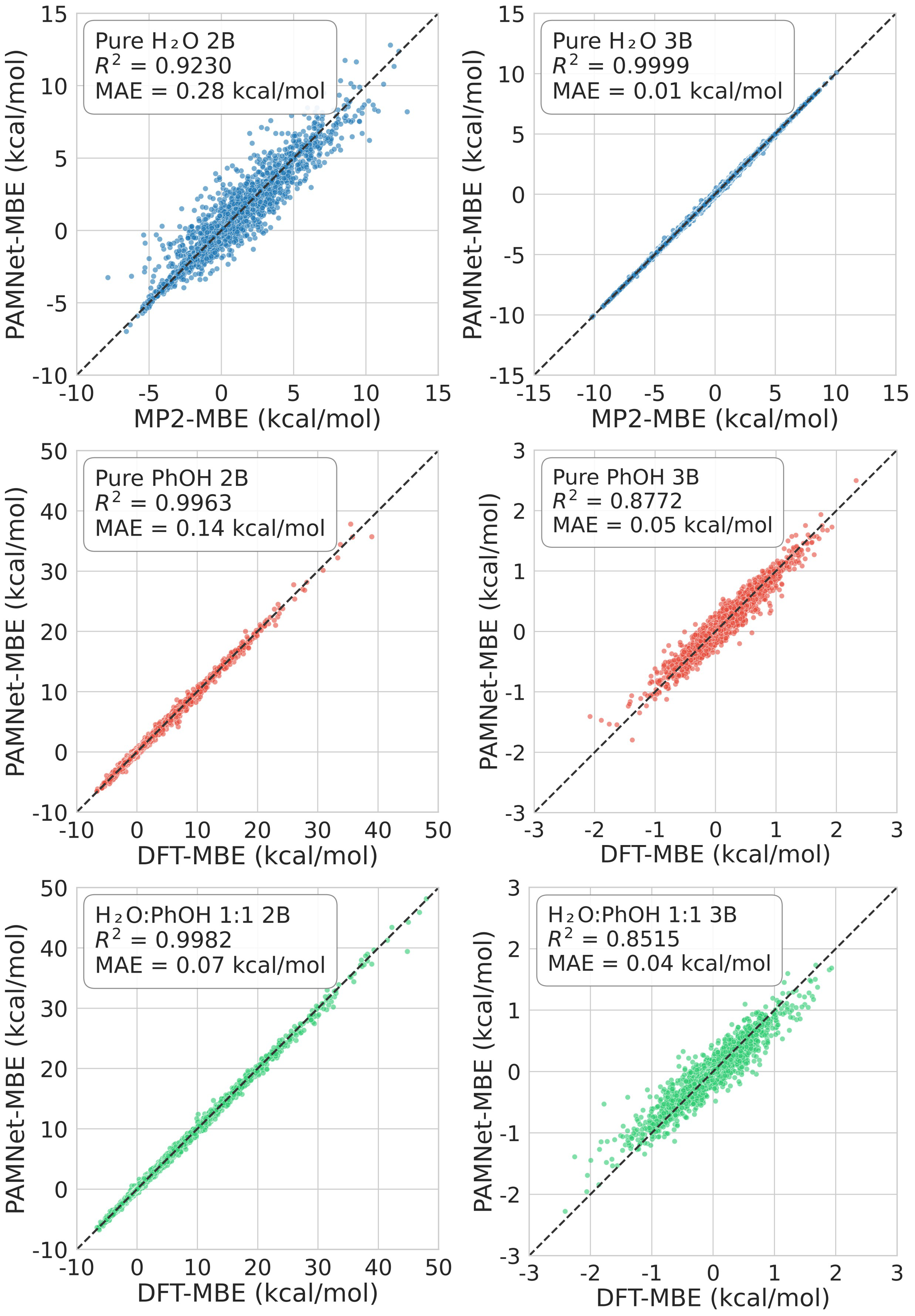}
\caption{2B (left) and 3B (right) energies on double-density water (top), phenol (middle), and 1:1 water:phenol (bottom) clusters are predicted by PAMNet-MBE using the original training strategy and compared with MP2-MBE or DFT-MBE.} 
\label{fig:MXMNet_PAMNet_results}
\end{figure}

To provide a proof-of-concept for 
the FB-GNN-MBE models (MXMNet-MBE and PAMNet-MBE), 
we will demonstrate that FB-GNN-MBE can reliably reproduce first-principles 2B and 3B energies across double-density benchmark systems of water, phenol, and 1:1 water:phenol clusters. 
We summarize the comparison between values calculated by MP2-MBE or DFT-MBE and those predicted by MXMNet-MBE or PAMNet-MBE in Figure \ref{fig:MXMNet_PAMNet_results} as well as in Figures S1 and S2 and Table S1 in the SM. 
MXMNet-MBE and PAMNet-MBE both demonstrate extremely high accuracy and efficiency, with high values of $R^2$, negligible mean (signed) errors (MEs), low mean absolute errors (MAEs), and reduced computational costs ($t_\text{rel} = \braket{t_\text{FB-GNN}}/\braket{t_\text{MP2/DFT}}$\textcolor{black}{, where $t_\text{FB-GNN}$ is the end-to-end wall time for each cluster, 
including graph construction, feature generation, and GNN 
inference, divided by the total number of 2B and 3B energies, and $t_\text{MP2/DFT}$ is the CPU time to calculate each dimer or trimer energy using MP2 or DFT).} 
These results indicate that FB-GNN-MBE  meets the chemical accuracy threshold and is reliable for replacing first-principles QM methods such as MP2 or DFT for low-order MBE corrections, with negligible computational cost and negligible systematic errors.\cite{doi:10.1021/jp211997b,yuk2024putting} 
Due to the doubled ($2.0\times$) densities of all these clusters, the 
high-energy repulsive regions of a PES are more frequently visited, so that FB-GNN-MBE learns a more comprehensive description of both the short- and long-range interactions in the 2B energies even when the distributions of values shift toward the positive ends.\cite{mahadevi2016cooperativity,grabowski2011covalency} 
3B energies are generally more challenging to model because they are typically smaller in magnitude due to the coexistence of cooperative and anti-cooperative effects.\cite{mahadevi2016cooperativity,grabowski2011covalency}
In addition, phenol and water:phenol clusters are more difficult to predict than water clusters because their intermolecular interactions are more complicated when $\pi$-electrons are involved, such as the increased electron delocalization, electron polarization, and $\pi$--$\pi$ stacking.\cite{kennedy2014communication,rezac2015benchmark,ma1997cation,liu2017capturing}
MXMNet-MBE and PAMNet-MBE exhibit marginally different behavior across these systems, suggesting that their performance is governed by underlying data quality and molecular physics rather than the specific model architecture.
However, the parallel message passing and attention fusion modules in PAMNet make it more flexible and interpretable 
than the multiplex architecture of MXMNet. 
Therefore, we present the results of PAMNet here in the main text and select it as the preferred FB-GNN backbone for future tasks.

\subsection{Comparison with Non-Fragment-Based Graph Neural Networks}

\begin{figure}[!h]
\centering
\includegraphics[width=0.5\textwidth]{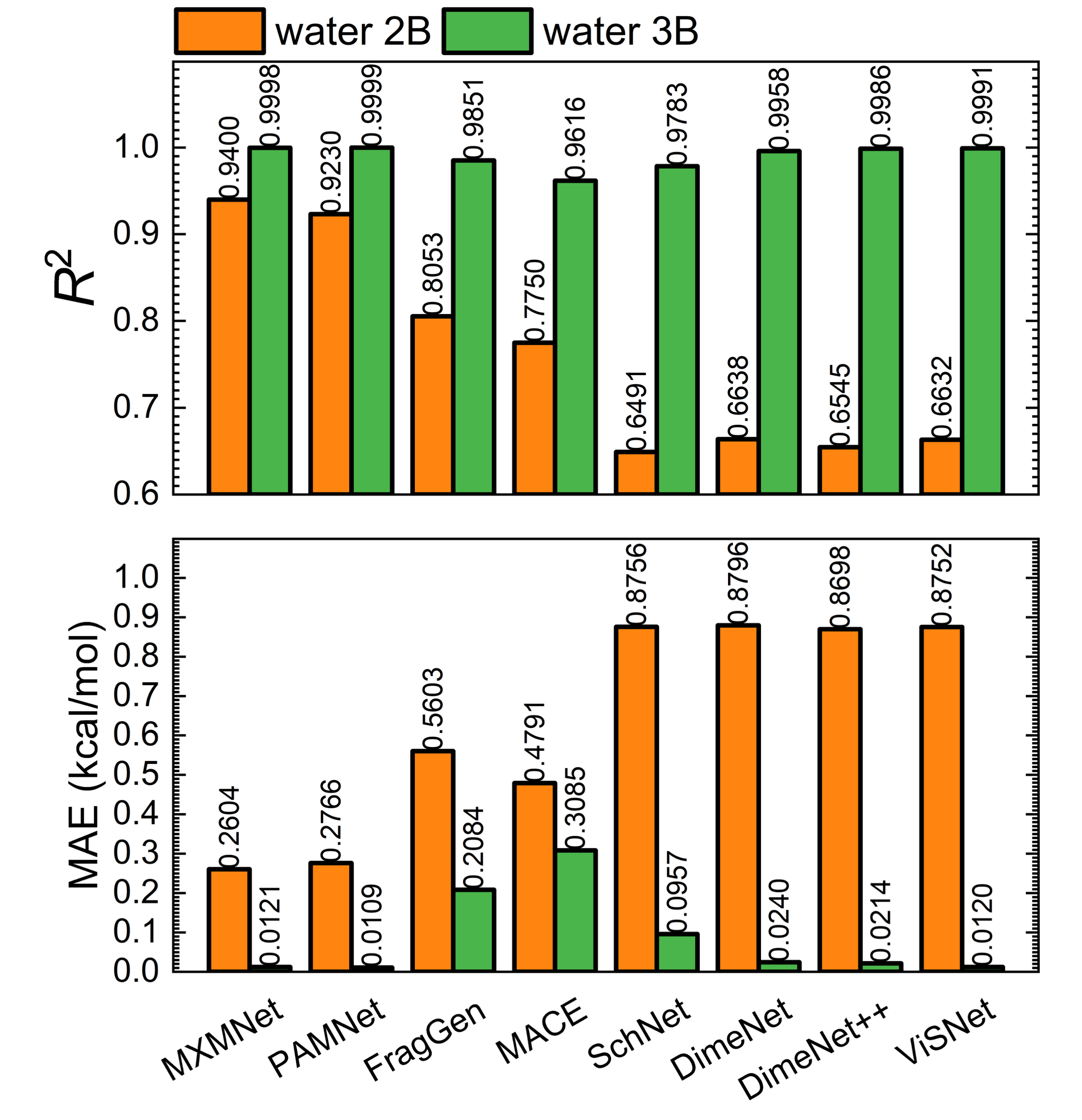}
\caption{Performance metrics of 2B and 3B energies on the double-density $\ch{(H2O)}_{67}$ clusters are compared among two FB-GNN-MBE methods and six non-FB-GNN-MBE methods using the original training strategy, in terms of $R^2$ (top) and MAE (kcal/mol, bottom).} 
\label{fig:comp_gnns}
\end{figure}

To validate the necessity of the fragment-based treatment in FB-GNN-MBE, we will demonstrate that FB-GNN-MBE is superior to GNN-MBE frameworks established using popular non-FB-GNNs, including MACE,\cite{batatia2022mace} SchNet,\cite{schutt2018schnet} DimeNet,\cite{gasteiger2020directional} DimeNet++,\cite{gasteiger2020fast} and ViSNet,\cite{wang2024enhancing} in predicting 2B and 3B energies from (\ch{H2O})$_{67}$ clusters.
We also compared with a transformer-based architecture of FragGen.\cite{zhang2024fraggen} 
As shown in Figure \ref{fig:comp_gnns} and Table S2 in the SM, MXMNet-MBE and PAMNet-MBE give the highest accuracy in both 2B and 3B energies.
MACE and FragGen achieve second-tier performance for 2B energies but are less accurate for 3B energies, while SchNet, DimeNet, DimeNet++, and ViSNet, show better performance in 3B energies than in 2B energies.
The strongest performance of MXMNet-MBE and PAMNet-MBE suggests that while directional and equivariant message passing in non-FB-GNN approaches are sufficient to model weaker, shorter-range interactions like the 3B energies, an explicit combination of short- and long-range geometric information implemented in the hierarchic architectures of FB-GNN approaches is more capable of capturing stronger, steeper, and mixed-range 2B energies effectively. 

\subsection{Analysis of Multi-Stage Training Strategy}
\label{sec:multi-stage-training}

\begin{figure}[!h]
\centering
\includegraphics[width=0.5\textwidth]{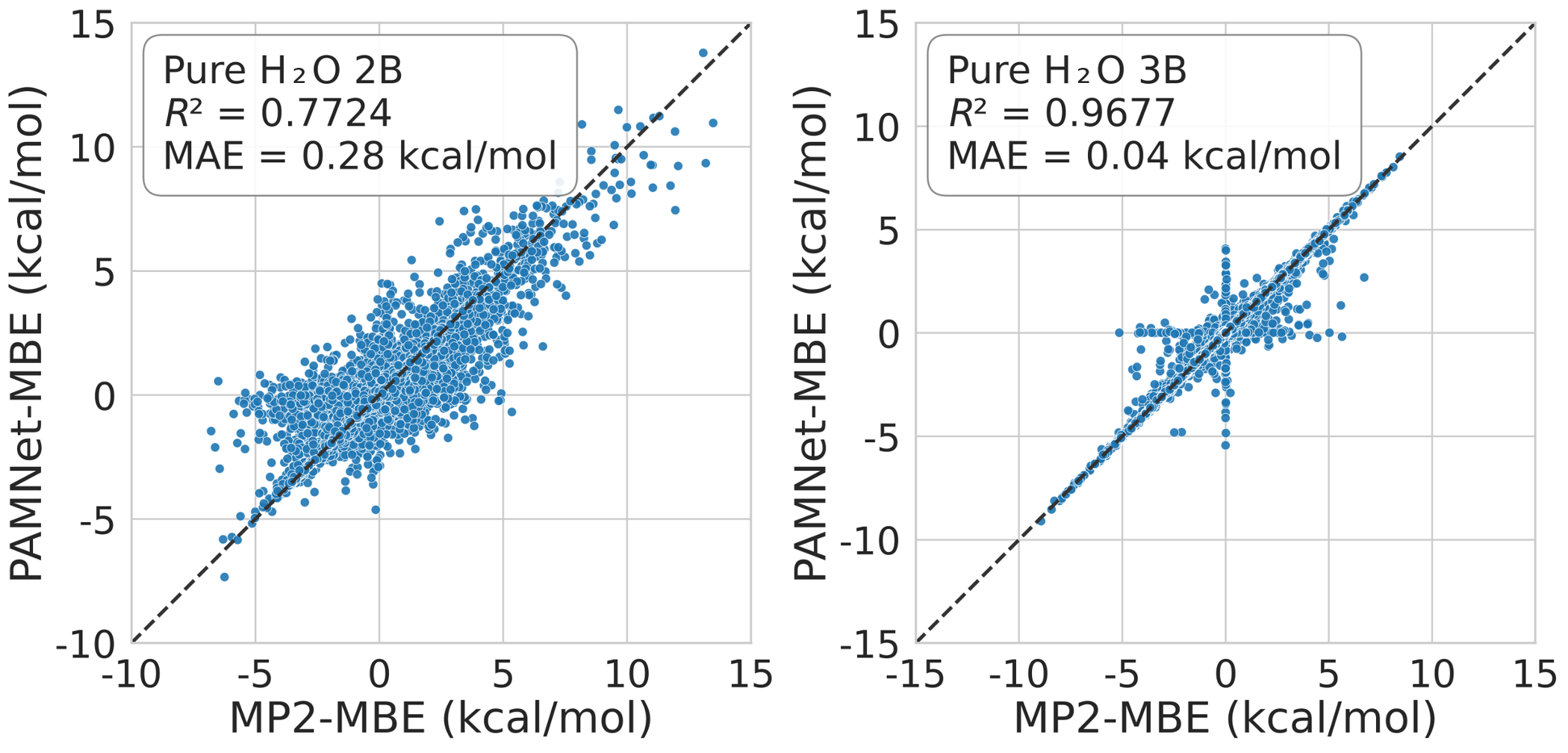}
\caption{2B (left) and 3B (right) energies on mixed-density water clusters are predicted by PAMNet-MBE using the multi-stage training strategy and compared with MP2-MBE.} 
\label{fig:multi_stage_results}
\end{figure}

To validate the implementation of the multi-stage training strategy,\cite{bengio2009curriculum} we will demonstrate that it is necessary when transferring from double-density clusters to mixed-density ones. 
This strategy was motivated by the need to cover molecular clusters under different environments (temperature, pressure, \textit{etc.}) and different phases (solid, liquid, \textit{etc.}) and the observation that the PAMNet-MBE model trained using the double-density significantly drops its performance on the mixed-density water clusters, showing low $R^2$ values of 0.6402 for 2B energies and 0.4612 for 3B energies.
Such a lack of transferability stems from the difficulty in fitting 
energy landscapes with different topological characters within a single training attempt, consistent with earlier observations.
\cite{kang2019decoupling,kim2023local,lai2024echomen} 
In addition, 3B energies face an additional challenge 
due to the 
predominance of near-zero values ($|E^\text{3B}_{ijk}|< 0.3$ kcal/mol in $\sim 78\%$ of samples).
As a result, the training algorithm may take a shortcut 
and predict zero values for all samples rather than learning the distinct repulsive and attractive features of the PES. 
As demonstrated in Figure \ref{fig:multi_stage_results}, as well as Figures S3 and S4 and Table S4 in the SM, our multi-stage training strategy successfully restores the exceptional model performance in 2B and 3B energies in phenol and mixture clusters with $2.0\times$ densities.
For the mixed-density water clusters, the 2B and 3B energies remain highly competitive, except that the $R^2$ values of 0.7724 and 0.9677 appear lower than the double-density cluster due to the statistical artifact from the expanded range of sampled configurations.
This result suggests that the core molecular physics underlying  intermolecular interactions is accurately captured by the multi-stage training strategy, which addresses 
the problem of imbalanced datasets by prioritizing high-energy repulsive and near-equilibrium regions of the PES before refining 
near-dissociation regions.\cite{mahadevi2016cooperativity} %

\subsection{
Analysis of One-Dimensional Potential Energy Surfaces}
\label{sec:pes}

\begin{figure}[!h]
\centering
\includegraphics[width=0.5\textwidth]{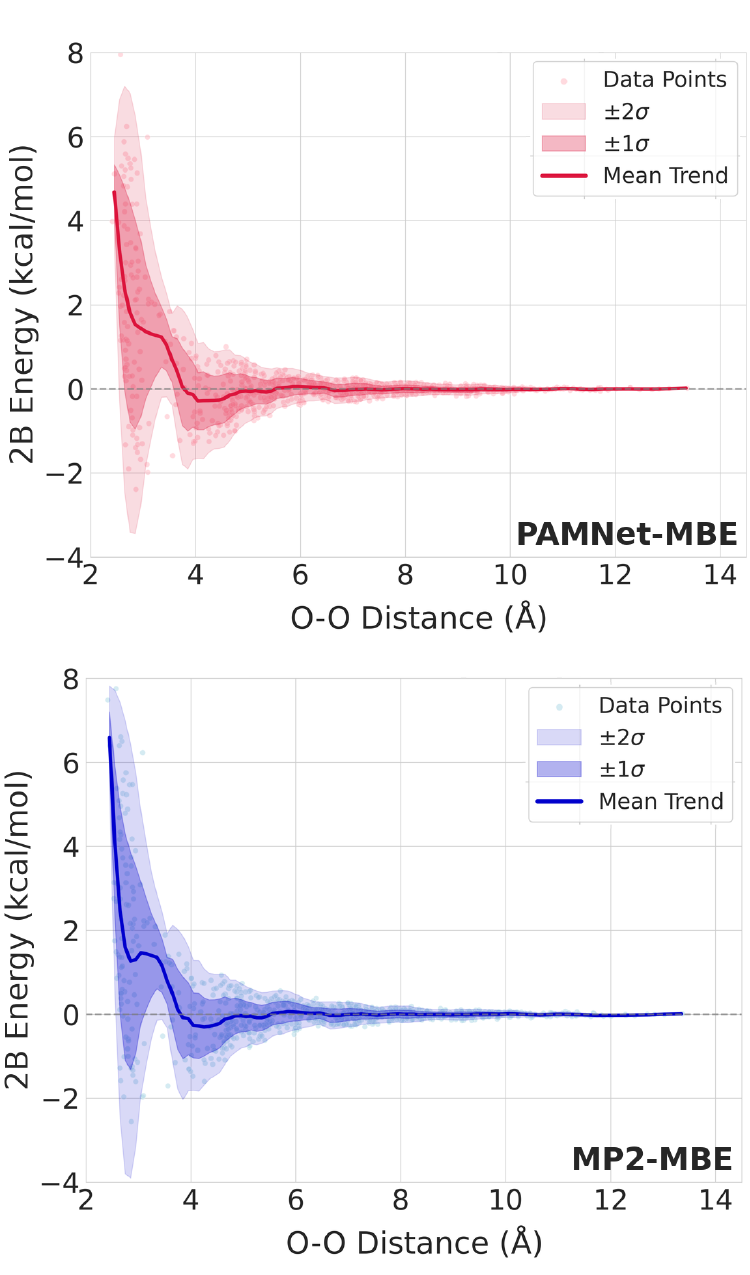}
\caption{Collection of 1D dissociation curves of all possible water dimers in a double-density $(\ch{H2O})_{67}$ cluster as functions of \ch{O-O} distance is compared between fine-tuned PAMNet-MBE (top) and MP2-MBE (bottom).} %
\label{fig:2b_pes_distribution_water}
\end{figure}



\begin{figure}[!ht]
\centering
\includegraphics[width=0.5\textwidth]{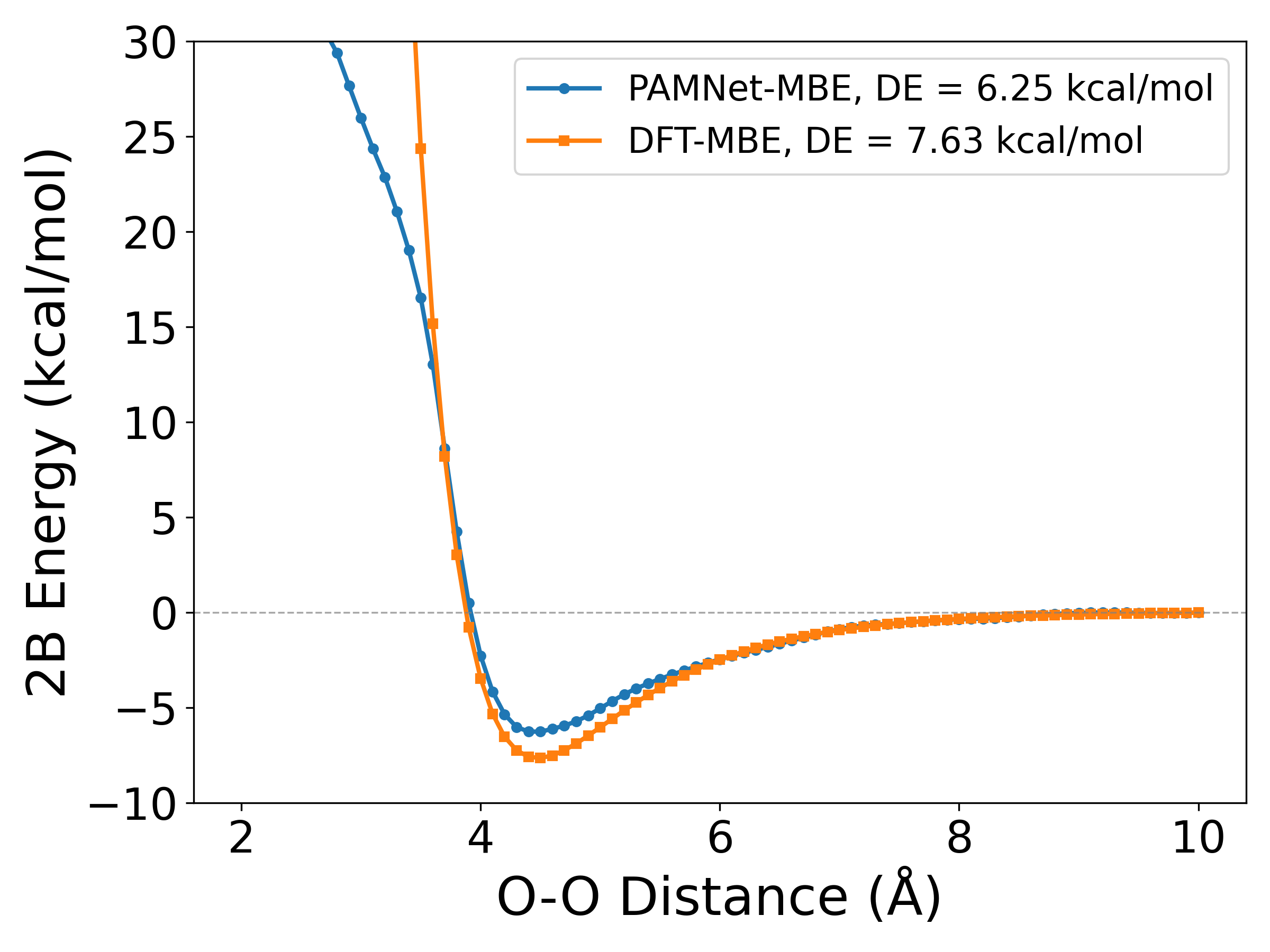}
\caption{1D dissociation curve of 
a random phenol dimer as a function of \ch{O-O} distance is compared between PAMNet-MBE and DFT-MBE, with dissociation energies (DE) labeled.} 
\label{fig:2b_pes_distribution_phenol}
\end{figure}

To further validate the generalizability of PAMNet-MBE 
over a full range of configurations, as if in a practical setting, we will demonstrate its ability to reconstruct the shape and value of the 1D PES (dimeric dissociation curve) from a double-density $(\ch{H2O})_{67}$ cluster (with a brief fine-tuning before inference) and two random phenol molecules (without any fine-tuning before inference). 
For both systems, we used the \ch{O-O} interatomic distance as 
the reaction coordinate and the primary variable that directly governs the strength of hydrogen bonds and van der Waals interactions.\cite{manchev2024modeling,abella2023many}
Figure \ref{fig:2b_pes_distribution_water} shows that the regular fine-tuning treatment of the original PAMNet-MBE model 
successfully generates the 1D PES of a water cluster, 
including a repulsive region with short O--O distances ($<2.8$ \AA), a near-equilibrium region with intermediate O--O distances ($2.8-5.0$ \AA), and a near-dissociation region with long O--O distances ($>5.0$ \AA).\cite{stone2013theory,soper2000structures}
The overall energy profile and its spread both well reproduce 
the MP2-MBE calculations. 
In particular, configurations in the near-equilibrium region 
exhibit a wide range of 2B energies, reflecting the significant impact of 
intermolecular orientations under a high density. 
Compared to the original 
PAMNet-MBE model (Table S4), the fine-tuned version exhibits a stronger performance, with a larger $R^2$ (0.9685 \textit{vs} 0.7960) and a smaller MAE (0.0807 \textit{vs} 0.2324 kcal/mol).
From Figure \ref{fig:2b_pes_distribution_phenol}, we observe a very similar behavior from 
a random phenol dimer. 
The dissociation energy is estimated to be 6.25 kcal/mol by fine-tuned PAMNet-MBE, which is close to DFT-MBE 
(7.63 kcal/mol) and the CCSD(T)/CBS-calculated value (6.84 kcal/mol) from Kim and coworkers. 
\cite{C003008B}
\textcolor{black}{This numerical agreement is surprising but likely due to 
error cancellation or representational smoothing. Achieving CCSD(T)-quality accuracy will require training and validation against CCSD(T) reference data.}
This result 
indicates that 
the fine-tuned PAMNet-MBE 
model successfully captures $\pi$--$\pi$ stacking interactions 
in phenol clusters, whereas highly directional and polarized hydrogen bonds in water clusters require more careful treatment. 

\subsection{Analysis of Teacher--Student Knowledge Distillation Protocol} 
\label{sec:fine-tune}

\begin{figure}[!h]
\centering
\includegraphics[width=0.5\textwidth]{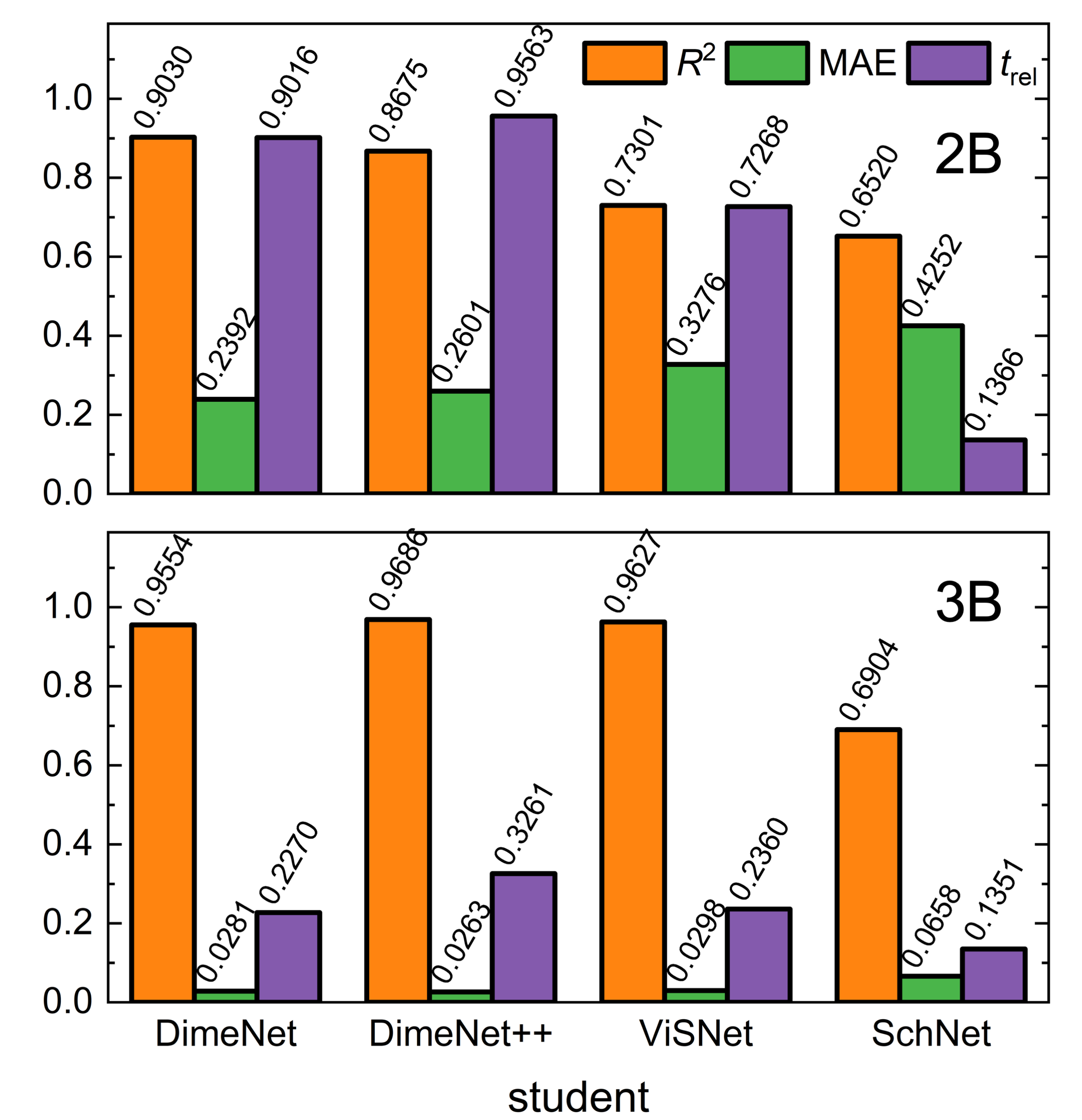}
\caption{Performance metrics of 2B (top) and 3B (bottom) energies on normal-density $\ch{(H2O)}_{21}$ clusters are compared among four fine-tuned student models (DimeNet-MBE, DimeNet++-MBE, ViSNet-MBE, and SchNet-MBE) using the teacher--student knowledge distillation protocol, in terms of $R^2$, MAE (kcal/mol), and $\braket{t_\text{rel}}$.} 
\label{fig:finetune_best}
\end{figure}

To validate the implementation of the teacher--student knowledge distillation protocol, we will demonstrate its generalizability to a small fine-tuning dataset of normal-density $\ch{(H2O)}_{21}$ clusters. 
Figure \ref{fig:finetune_best} and Table S5 in the SM summarize the performance of four selected student models, DimeNet-MBE, DimeNet++-MBE, ViSNet-MBE, and SchNet-MBE. 
DimeNet-MBE and DimeNet++-MBE achieve the best performance for both 2B and 3B energies but incur the longest inference times, reflecting that the explicit angular terms provide reasonable descriptions of the dimeric and trimeric interactions at the cost of increased computational complexity. 
ViSNet-MBE achieves a slightly lower accuracy with a comparable computational cost, suggesting that its equivariant vector features are also well-suited for electronic structures in non-covalent interactions. 
SchNet-MBE is much faster than other models but produces the largest errors, indicating that its distance-only representations are insufficient for angle-dependent non-covalent interactions. 
These results set a crucial stage for the application of the teacher--student protocol, because they show a successful information flow from the teacher to the student using a limited fine-tuning dataset with a very different distribution from the original training set. 

\begin{figure}[!ht]
\centering
\includegraphics[width=0.5\textwidth]{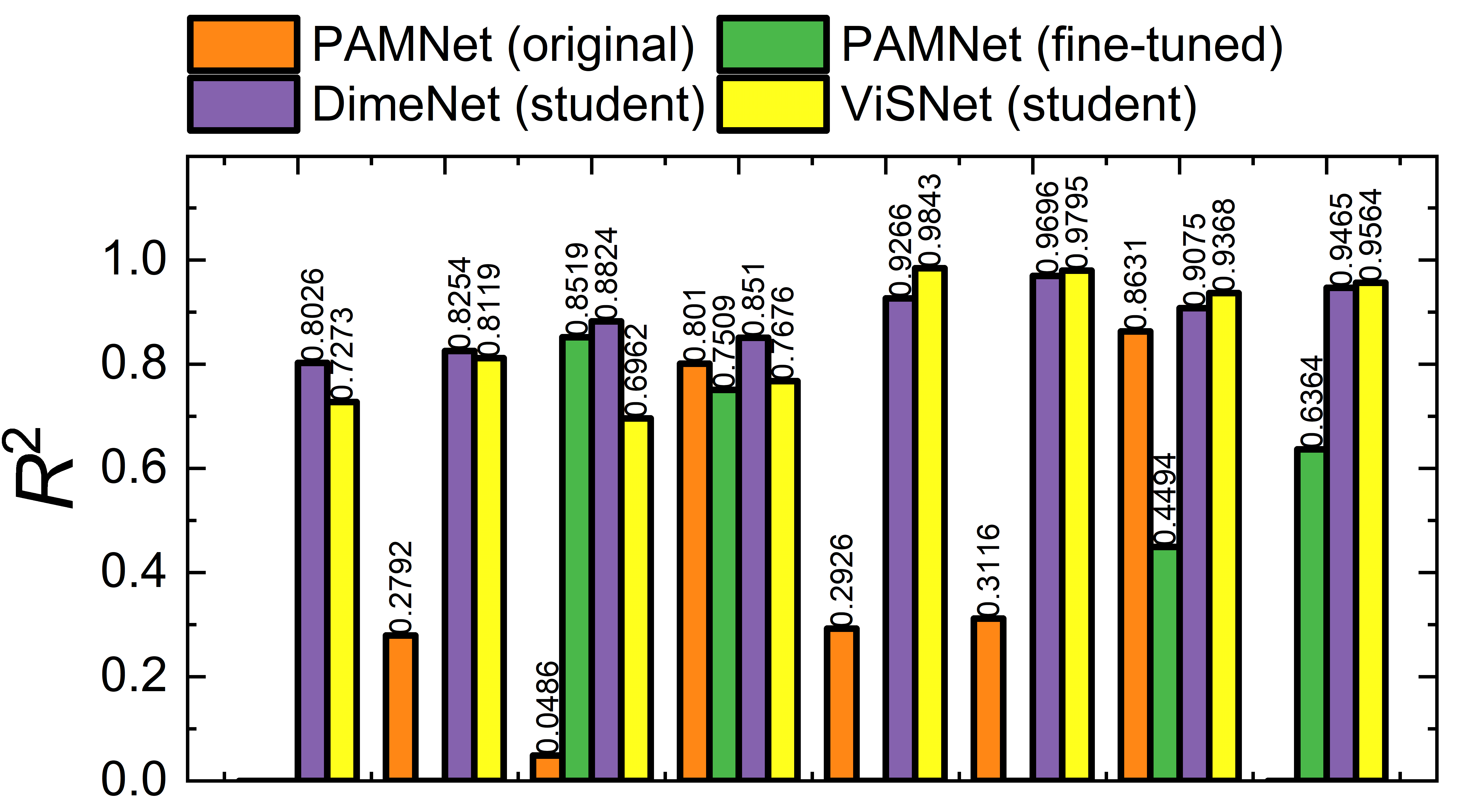}
\includegraphics[width=0.5\textwidth]{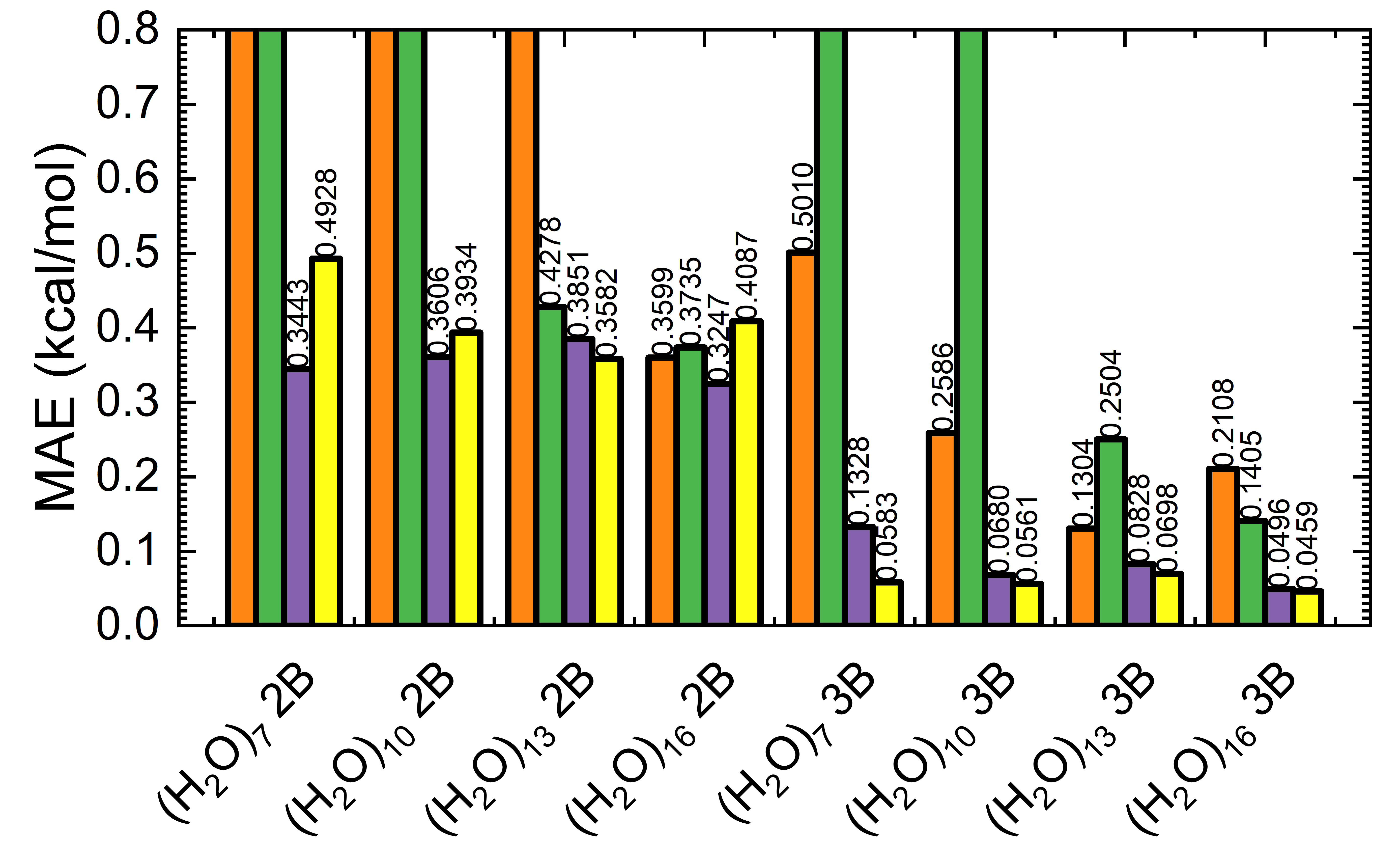} 
\caption{Performance metrics of 2B and 3B energies on normal-density small water clusters are compared among the original and fine-tuned teacher models (PAMNet-MBE), 
and the fine-tuned student models (DimeNet-MBE and ViSNet-MBE) 
using the teacher--student knowledge distillation protocol, in terms of $R^2$ (top) and MAE (kcal/mol, bottom). Data with $R^2<0$ and MAE $>0.8$ kcal/mol are not present in the scale.} 
\label{fig:transfer_all_models_2B3B}
\end{figure}

To further validate the generalizability of the teacher--student 
protocol, we will demonstrate the success of the fine-tuned, well-performing DimeNet-MBE and ViSNet-MBE models on four smaller, completely unseen water clusters, $\ch{(H2O)_{7}}$, $\ch{(H2O)_{10}}$, $\ch{(H2O)_{13}}$, and $\ch{(H2O)_{16}}$, without 
additional retraining. 
We evaluated the performance of DimeNet-MBE and ViSNet-MBE  
in 2B and 3B energies and compared them with the teacher PAMNet-MBE models that were trained from scratch (original) and regularly fine-tuned (fine-tuned), 
and summarize their results in Figures \ref{fig:transfer_all_models_2B3B}, as well as Figure S5 and Tables S6--S8 in the SM.
DimeNet-MBE trained using the teacher--student 
protocol consistently achieves low MAE values and high $R^2$ values across all unseen clusters, maintaining strong agreement with MP2-MBE 
over the full energy range.
DimeNet-MBE also captures a pronounced proportion of near-zero values, consistent with the reduced number of hydrogen bonds in small clusters.\cite{herman2023extensive} 
{From a modeling perspective, this behavior reemphasizes the importance of the angular features of DimeNet in modeling short-range, geometry-resolved interactions, which also vanish for distant dimers 
to avoid spurious variance in 1D PES.} 
In addition, the knowledge distillation treatment dramatically improves performance 
relative to the original and fine-tuned PAMNet-MBE models, but the advantage decays as the cluster size approaches that from 
the original training set. 
An extreme case is the 3B energies of $\ch{(H2O)_{7}}$, 
for which the MAE of the original PAMNet-MBE is $8.6$ times that of ViSNet-MBE (0.5010 \textit{vs} 0.0583 kcal/mol). 
The standard fine-tuning strategy used to construct the 1D PES also fails in this regime, yielding a negative $R^2$ value, 
confirming that naive parameter transfer without knowledge distillation destroys pre-trained representations when target data are scarce.

As an important sanity check of our teacher--student 
protocol and the fragmentation strategy, we evaluated the total ground state energies of these four small water clusters using fine-tuned DimeNet-MBE and ViSNet-MBE, as in Equation (\ref{eq:mbe}), and present results in Table \ref{tab:transfer_all_models_full}. 
Both models reproduce 
the results from MP2-MBE and non-MBE MP2 calculations with negligible errors, demonstrating high fidelity and small truncation errors.\cite{heindel2023many,riera2023mbx}
The success of the student models on unseen water clusters, even after limited fine-tuning, confirms that the teacher--student 
protocol is able to capture the molecular physics from MBE and transfer this knowledge from a general domain to a specific one, offering a universal and 
efficient surrogate to reproduce high-fidelity PES calculations. 

\begin{table}[!ht]
\centering
\caption{Total ground state energies of normal-density small water clusters are predicted by DimeNet-MBE and ViSNet-MBE using the teacher--student knowledge distillation protocol and compared with MP2-MBE and MP2 benchmarks, along with percentage errors ($\Delta$).} 
\begin{tabular}{c|c|c|c|c|c|c|c}
\hline\hline
cluster & \multicolumn{2}{c|}{DimeNet-MBE} & \multicolumn{2}{c|}{ViSNet-MBE} & \multicolumn{2}{c|}{MP2-MBE} & MP2 \\
\cline{2-8}
& $E$ (hartree) & $\Delta$ (\%) & $E$ (hartree) & $\Delta$ (\%) & $E$ (hartree) & $\Delta$ (\%) & $E$ (hartree)\\
\hline\hline
$\ch{(H2O)}_{7}$  & $-533.8904$ & 0.0062 & $-533.8943$ & 0.0055 & $-533.8916$ & 0.0060 & $-533.9235$ \\ 
\hline
$\ch{(H2O)}_{10}$ & $-762.7071$ & 0.0079 & $-762.7092$ & 0.0076 & $-762.7148$ & 0.0069 & $-762.7672$ \\ 
\hline
$\ch{(H2O)}_{13}$ & $-991.5215$ & 0.0087 & $-991.5437$ & 0.0065 & $-991.5350$ & 0.0074 & $-991.6080$ \\ 
\hline
$\ch{(H2O)}_{16}$ & $-1220.3182$ & 0.0109 & $-1220.3218$ & 0.0106 & $-1220.3583$ & 0.0076 & $-1220.4513$ \\ 
\hline\hline
average  &  & {0.0084} & & {0.0075} & & {0.0070} &  \\ 
\hline\hline
\end{tabular}%
\label{tab:transfer_all_models_full}
\end{table}

\section{Conclusion and Outlook}
\label{sec:conclusion}

In the present study, we introduced FB-GNN-MBE, a computational framework that couples FB-GNNs with the MBE theory and delivers robustness, accuracy, and interpretability, thereby outperforming conventional NN- or GNN-accelerated QM models for large chemical systems. 
FB-GNN-MBE overcomes the unfriendly computational complexity 
by computing only 1B energies at the QM level while predicting many-body corrections (including 2B and 3B energies) based on FB-GNN-trained structure--property relationships. 
Using 
MXMNet\cite{zhang2020molecular} and PAMNet\cite{zhang2023universal} as the backbone FB-GNN models, we improve the alignment between the model architecture and the chemical hierarchy. 
Across water, phenol, and 1:1 water:phenol clusters, FB-GNN-MBE achieves chemical accuracy in ground state energies in agreement with MP2-MBE or DFT-MBE, while reducing the computational cost by {two to four} orders of magnitude.
In addition, we presented a practical teacher--student knowledge distillation protocol for building 
transferable FB-GNN-MBE models without system-specific 
re-training 
and for removing the central bottleneck of overfitting in machine learned force fields (MLFFs).
We showed that the chemical patterns learned by a large, heavy-weight teacher model (such as PAMNet) can be distilled and transferred to multiple small, light-weight student models (such as DimeNet and ViSNet).
In summary, 
FB‑GNN‑MBE 
is an accurate and scalable approach for 
\textcolor{black}{predicting interaction energies of large molecular assemblies and is readily applicable to MC sampling or any other energy-based screening workflow.}\cite{wales1997global,laio2002escaping,paesani2016getting}
FB-GNN-MBE exhibits the following strengths: 
(1) It is robust to data imbalance between low- and high-energy regions of the PES using the multi-stage training strategy. 
(2) It is generalizable to under-sampled regions of the PES using the knowledge distillation and fine-tuning strategies.
(3) Its model architecture encodes explicit directional information using distance- and angle-based features. 

For future studies, we plan to transfer FB-GNN-MBE from non-covalently bonded clusters to covalently or ionically bonded systems with stronger interfragment interactions and a bond-cleaving fragmentation scheme.\cite{giese2017quantum,wang2014quantum}
\textcolor{black}{It is worth noting that for such a system fragmentation strategy is less well-defined than a non-covalent cluster discussed in the present study because a single molecule is no longer a natural fragment. 
As a result, we expect to consider additional factors, such as 
capping atoms/groups, charge/spin assignment, and convergence test.\cite{10.1063/1.5126216,ruther2026comprehensive}
We will discuss part of the fragmentation strategy in Section II of SM.
In a future architecture of FB-GNN-MBE, we will 
introduce features describing fragment boundaries and semi-empirical electronic structures to complement the current geometry-only representation.}\cite{parsaeifard2021fingerprint} 

We also plan to introduce two complementary forms of supervision through a multi-task head \textcolor{black}{so that multiple many-body properties can be evaluated in addition to the energies as regularizers in the loss functions}.\cite{caruana1993multitask,unke2019physnet,XU202247}
\textcolor{black}{First, we will perform} the joint energy--force training.\cite{behler2007generalized} 
\textcolor{black}{Although the present study focuses on energies, atomic forces are equally critical for large-scale molecular simulations because they shape the curvature of the PES. 
1B and $n$B contributions of atomic forces can be evaluated as the analytical gradient of Equation (\ref{eq:1b}) and trainable numerical gradients of Equations (\ref{eq:2b}) and (\ref{eq:3b}), respectively. 
Second, we will perform} the energy decomposition analysis (EDA).\cite{mao2017energy}
\textcolor{black}{We will break down the interfragment interactions into trainable frozen, polarization, charge transfer, and dispersion interactions, and improve} the physical interpretation of \textcolor{black}{FB-GNN-MBE} 
and 
disentangles overlapping electronic effects in 
polarizable systems.\cite{wang2014quantum}

\section*{Author Contributions}
Siqi Chen and Zhiqiang Wang contributed equally to this work. Siqi Chen: Data curation, Formal analysis, Investigation, Methodology, Software, Validation, Visualization, Writing--original draft, Writing--review \& editing. Zhiqiang Wang: Data curation, Formal analysis, Methodology, Software, Visualization. Yili Shen: Formal analysis, Methodology, Software, Visualization. Xianqi Deng: Formal analysis, Methodology, Software, Validation, Visualization. Xi Cheng: Data curation, Formal analysis. Cheng-Wei Ju: Conceptualization, Formal analysis, Investigation, Methodology, Validation. Jun Yi: Data curation. Guo Ling: Data curation. Dieaa Alhmoud: Data curation. Hui Guan: Conceptualization, Funding acquisition, Investigation, Methodology, Project administration, Resources, Software, Supervision, Validation. Zhou Lin: Conceptualization, Data curation, Formal analysis, Funding acquisition, Investigation, Methodology, Project administration, Resources, Software, Supervision, Validation, Visualization, Writing--original draft, Writing--review \& editing.

\begin{acknowledgments}
	Z.L. and H.G. thank the financial support from UMass Amherst Start-Up Funds,  
	UMass ADVANCE Collaborative Research Seed Grant,
	and National Science Foundation (\#CISE/IIS-2435822).
	S.C. and X.C. thank the financial support from the PPG Fellowships.
	All authors thank fruitful discussions with Prof. Lei Xie, Dr. Shuo Zhang, Dr. You Wu, Prof. Joel Bowman, and Dr. Kun Yao, as well as high-performance supercomputing resources provided by UMass/URI Unity Cluster and MIT Supercloud.\cite{8547629}
\end{acknowledgments}

\section*{Supplementary Material}
{Supplementary Material provides (1) implementation and validation details for the FB-GNN-MBE framework, including the source code availability and pseudo-algorithms for MXMNet-MBE, PAMNet-MBE, multi-stage training, and teacher--student knowledge distillation; (2) fragmentation strategies for the present and future studies; (3) mathematical definitions of evaluation metrics, including $R^2$, ME, MAE, and MSE; (4) additional performance assessment, including benchmarks of MXMNet-MBE and/or PAMNet-MBE on double-density clusters, mixed-density clusters, and 1D PES, and benchmarks of light-weight fine-tuned student models; (5) UMAP analysis that visualizes learned 2B latent spaces; and (6) hyperparameter tables and sensitivity analysis about teacher--student knowledge distillation. 
}

\clearpage
\begin{center}
\Large \textbf{SUPPLEMENTARY MATERIAL}
\end{center}

\section{Source Code and Pseudo-Algorithms}
\label{sec:alg}

\subsection{Online Repository}
We uploaded our source code 
associated with our FB-GNN-MBE models for training, validation, and testing 
to GitHub ({\url{https://github.com/Lin-Group-at-UMass/FBGNN-MBE}}).
We also uploaded all datasets to FigShare ({\url{https://doi.org/10.6084/m9.figshare.31933827}}).

\subsection{Pseudo-Algorithms}
To enable readers to reproduce our training results for two versions of FB-GNN-MBE, the multi-stage training strategy, and the teacher--student knowledge distillation framework, 
we provide the protocols below as pseudo-algorithms.


\clearpage

\begin{algorithm}[H]
\SetAlgoLined
\DontPrintSemicolon
\KwIn{Molecular graph: atom types $x$, 3D coordinates $\text{pos}$, edge indices, batch indices}
\KwOut{Predicted n-body interaction energy $\hat{E}$}
\caption{MXMNet-MBE}

\textbf{Step 1: Embedding and graph construction}\;
$h \gets \text{MLP}(\text{Embedding}(x))$\;
$(\mathcal{E}_g, d_g) \gets \text{RadiusGraph}(\text{pos}, r_g)$\;
$(\mathcal{E}_l, d_l) \gets \text{RemoveSelfLoops}(\text{edge\_index})$\;
${I} \gets \text{ComputeTripletIndices}(\mathcal{E}_l)$\;

\textbf{Step 2: Geometric feature encoding}\;
$\text{rbf}_g \gets \text{MLP}(\text{BesselBasis}(d_g))$\; 
$\text{rbf}_l \gets \text{MLP}(\text{BesselBasis}(d_l))$\;
$\theta_1 \gets \text{Angles}(\text{pos}, \mathcal{I}_{\text{one-hop}})$\; 
$\theta_2 \gets \text{Angles}(\text{pos}, \mathcal{I}_{\text{two-hop}})$\;
$\text{sbf}_1 \gets \text{MLP}(\text{SphericalBasis}(d_l, \theta_1))$\; $\text{sbf}_2 \gets \text{MLP}(\text{SphericalBasis}(d_l, \theta_2))$\;

\textbf{Step 3: Hierarchical message passing}\;
\For{$\ell = 1$ \KwTo $L$}{
    $h \gets \text{GlobalMP}^{(\ell)}(h, \text{rbf}_g, \mathcal{E}_g)$\;
    $h \gets \text{LayerNorm}(h)$ \;
    $h \gets \text{LocalMP}^{(\ell)}(h, \text{rbf}_l, \text{sbf}_1, \text{sbf}_2, {I}, \mathcal{E}_l)$\;
}

\textbf{Step 4: Readout and prediction}\;
$h_{\text{graph}} \gets \text{GlobalAddPool}(h, \text{batch})$ \;
$\hat{E} \gets \text{MLP}(h_{\text{graph}})$\;
\Return{$\hat{E}$}
\end{algorithm}

\begin{algorithm}[H]
\SetAlgoLined
\DontPrintSemicolon
\KwIn{Molecular graph: atom types $x$, 3D coordinates $\text{pos}$, edge indices, batch indices}
\KwOut{Predicted n-body interaction energy $\hat{E}$}
\caption{PAMNet-MBE}

\textbf{Step 1: Embedding and graph construction}\;
$h \gets \text{MLP}(\text{Embedding}(x))$\; 
$(\mathcal{E}_g, d_g) \gets \text{RadiusGraph}(\text{pos}, r_g)$\; 
$(\mathcal{E}_l, d_l) \gets \text{RemoveSelfLoops}(\text{edge\_index})$\; 
${I} \gets \text{ComputeTripletIndices}(\mathcal{E}_l)$

\textbf{Step 2: Geometric feature encoding}\;
$\text{rbf}_g \gets \text{MLP}(\text{BesselBasis}(d_g))$\; 
$\text{rbf}_l \gets \text{MLP}(\text{BesselBasis}(d_l))$\;
$\theta_1 \gets \text{Angles}(\text{pos}, \mathcal{I}_{\text{one-hop}})$\; 
$\theta_2 \gets \text{Angles}(\text{pos}, \mathcal{I}_{\text{two-hop}})$\;
$\text{sbf}_1 \gets \text{MLP}(\text{SphericalBasis}(d_l, \theta_1))$\; $\text{sbf}_2 \gets \text{MLP}(\text{SphericalBasis}(d_l, \theta_2))$\;

\textbf{Step 3: Hierarchical message passing}\;
$\mathbf{H} \gets []$  \\
\For{$\ell = 1$ \KwTo $L$}{
    $h \gets \text{GlobalMP}^{(\ell)}(h, \text{rbf}_g, \mathcal{E}_g)$\;
    $h \gets \text{LocalMP}^{(\ell)}(h, \text{rbf}_l, \text{sbf}_1, \text{sbf}_2, {I}, \mathcal{E}_l)$\;
    Append $\text{GlobalAddPool}(h, \text{batch})$ to $\mathbf{H}$\;
}

\textbf{Step 4: Attention-based multi-layer fusion}\;
$\mathbf{H}_{\text{stack}} \gets \text{Stack}(\mathbf{H})$ \\
$\alpha \gets \text{Softmax}(\text{LeakyReLU}(\text{Linear}(\mathbf{H}_{\text{stack}})))$ \\ 
$h_{\text{graph}} \gets \sum_{\ell}(\alpha_\ell \cdot \mathbf{H}_\ell)$ 

\textbf{Step 5: Prediction}\;
$\hat{E} \gets \text{Linear}(h_{\text{graph}})$\;
\Return{$\hat{E}$}
\end{algorithm}

\begin{algorithm}[H]
\SetAlgoLined
\DontPrintSemicolon
\KwIn{Model $M$, Dataset $D$, Mode $\in \{\text{`train'}, \text{`predict'}\}$}
\KwOut{Trained Model $M^*$ (if train) or Predictions $\hat{Y}$ (if predict)} 
\caption{Multi-Stage Training}
\uIf{mode == `train'}{
    Initialize $M$; Compute $y_{min}, y_{max} \Leftarrow \text{MinMax}(D.y)$\;
    Compute thresholds: $\tau_{high} \Leftarrow P_{75}(|E|)$, $\tau_{med} \Leftarrow P_{50}(|E|)$\;
    
    \textbf{Stage 1: High-energy-magnitude subset} ($|E| > \tau_{high}$)\\
    $D^{(1)} \Leftarrow \text{Filter}(D, |E| > \tau_{high})$ \\ 
    \For{epoch $e = 1$ \KwTo $E_1$}{
        Train $M$ on $D^{(1)}$ with $\mathcal{L} = \text{L1Loss}(M(x), y_{norm})$\;
    }
    Save best $M_1^*$\;
    
    \textbf{Stage 2: Medium-energy-magnitude subset} ($|E| > \tau_{med}$)\\
    Load $M \Leftarrow M_1^*$\;
    $D^{(2)} \Leftarrow \text{Filter}(D, |E| > \tau_{med})$ \\ 
    \For{epoch $e = 1$ \KwTo $E_2$}{
        Train $M$ on $D^{(2)}$ with reduced LR\;
    }
    Save best $M_2^*$\;
    
    \textbf{Stage 3: Full dataset}\\
    Load $M \Leftarrow M_2^*$\;
    \For{epoch $e = 1$ \KwTo $E_3$}{
        Train $M$ on full $D$ with lowest LR\;
    }
    Save $M^*$\;
}
\ElseIf{mode == `predict'}{
    Load $M^*$ and $(y_{min}, y_{max})$\;
    \For{batch $B \in D_{target}$}{
        $\hat{y} \Leftarrow \text{Denormalize}(M^*(B), y_{min}, y_{max})$\;
    }
    Compute MAE (kcal/mol), $R^2$\;
    \Return{Predictions $\hat{Y}$, Metrics}
}
\end{algorithm}

\begin{algorithm}[H]
\SetAlgoLined
\DontPrintSemicolon
\KwIn{Teacher $T$, Student $S$, Dataset $D$, Mode $\in \{\text{`train'}, \text{`predict'}\}$}
\KwOut{Trained Student $S^*$ (if train) or Predictions $\hat{Y}$ (if predict)} 
\caption{Teacher-Student Knowledge Distillation}
\uIf{mode == `train'}{
    Load Teacher $T$; Initialize Student $S$; Load Dataset $D$\;
    Compute: $y_{min}, y_{max} \Leftarrow \text{MinMax}(D.y)$\;
    
    \textbf{Stage 1: Knowledge distillation}\\
    \For{epoch $e = 1$ \KwTo $E_{distill}$}{
        \For{batch $B \in D$}{
            $y_T, f_T \Leftarrow T(B)$; \quad $y_S, f_S \Leftarrow S(B)$\;
            $L \Leftarrow \text{MSE}(y_S, y_T) + \lambda \cdot \text{MSE}(f_S, f_T)$\;
            Update $S$ via backpropagation\;
        }
    }
    
    \textbf{Stage 2: Fine-tuning on true labels}\\
    \For{epoch $e = 1$ \KwTo $E_{ft}$}{
        \For{batch $(x, y) \in D$}{
            $L \Leftarrow \text{L1Loss}(S(x), y_{norm})$\;
            Update $S$ via backpropagation\;
        }
    }
    Save $S^*$ and $(y_{min}, y_{max})$\;
}
\ElseIf{mode == `predict'}{
    Load Student $S^*$ and $(y_{min}, y_{max})$\;
    \For{batch $B \in D_{target}$}{
        $\hat{y}_{norm} \Leftarrow S^*(B)$\;
        $\hat{y} \Leftarrow \text{Denormalize}(\hat{y}_{norm}, y_{min}, y_{max})$\;
    }
    Compute MAE (kcal/mol), $R^2$\;
    \Return{Predictions $\hat{Y}$, Metrics}
}
\end{algorithm}

\section{{Fragmentation Strategies}}
\label{sec:strategy}

\textcolor{black}{At the current development stage of FB-GNN-MBE, 
the fragmentation strategy is determined case-by-case and depends on the chemical properties the systems in question.
For the current non-covalently bonded systems and future covalently and ionically bonded systems, we have so far implemented a ``top-down'' fragmentation strategy with two principles: 
(1) We maintain the smallest functionally meaningful unit. 
(2) 
We break only single bonds 
or non-bonding interactions.\cite{D0CP06528E}} 

\textcolor{black}{In the present proof-of-concept work we studied non-covalently bonded systems like water aggregates, phenol aggregates, and water--phenol mixtures. In these systems
every single molecule is a natural fragment and an intermolecular interaction like hydrogen bond and van der Waals force plays an essential role in $n$-body energies ($n\ge 2$). 
For our future studies about covalently bonded systems like organic polymers, we have tentatively planned to extend the fragmentation strategy, in which we will 
treat each monomer as a fragment, cleaving the single carbon--carbon (\ch{C-C}) bonds and capping any unsaturated bonds with a hydrogen atom (\ch{-H}) or a hydroxyl group (\ch{-OH}) depending on the effective fragment charge. 
We will try our best not to break any complete functional groups like phenyl (\ch{-C6H5}) and carboxylic acid (\ch{-COOH}) or known unit like an amino acid and a DNA base.
When we evaluate the many-body energies or forces, we will remove the contributions from capping atoms/groups as molecules (\ch{H2}, \ch{H2O}, \ch{H2O2}, \textit{etc.}}).

\textcolor{black}{For a new system, we will perform a brief convergence test for the fragmentation strategy. We will run a low-level first-principles MBE(2) calculation where the total energy is truncated at the two-fragment interactions to make sure the fragmentation does not introduce huge estimated error to damage the chemistry.
In this way the system will maintain the smallest functional or repeating unit such as a full $\pi$-conjugation, and our model can capture short-range and long-range interactions to simulate the real chemical systems, such as covalent bonds, ionic bonds, hydrogen bonds, London dispersions, dipole--dipole interactions, $\pi$--$\pi$ stacking, and solvation effects.}

\textcolor{black}{If a system does not exhibit natural or obvious fragments, such as polyacetylene and polytethlyene, we will consider 
an alternative ``bottom-up'' strategy, where we will construct fragments from individual functional groups, or monomers until some convergence is reached using a low-level first-principles MBE(2) calculation.
In this way we will allow fragments to grow around each unit and accommodate underlying effects from local electronic environment and molecular conformations.}

\textcolor{black}{We foresaw great room for us to explore and validate our fragmentation strategies and enhance our model versatility and accuracy in future studies. 
Table II in the main text can serve as a preliminary validation of 
our top-down fragmentation strategy. 
When we treated each water molecule as a single fragment, we reached an average of 0.0070\% 
relative errors for these water clusters using 
MP2-MBE(3), confirming a favorable choice of fragmentation strategy.}
\section{Evaluation Metrics}
\label{sec:def}

We set MP2- or DFT-evaluated 2B and 3B energies as $\{x_i\}$ and their FB-GNN-predicted counterparts as $\{y_i\}$.

\subsection{$\boldsymbol{R}$-Squared}

$R^2$ is the coefficient of determination between $\{x_i\}$ and $\{y_i\}$: 
\begin{equation}
    R^2 = 1- \dfrac{\displaystyle{\sum_{i=1}^N (y_i - x_i)^2}}{\displaystyle{\sum_i \left(x_i - \dfrac{1}{N}\sum_{i=1}^N x_i\right)^2}}
\end{equation}




\subsection{Mean (Signed) Error}
Mean (signed) error (ME) is defined as the average of the signed difference between $\{x_i\}$ and $\{y_i\}$: 
\begin{equation}
\text{ME}=\dfrac{1}{N}\sum_{i=1}^N(y_i - x_i)
\end{equation}

\subsection{Mean Absolute Error}
Mean absolute error (MAE) is defined as the average of the unsigned difference between $\{x_i\}$ and $\{y_i\}$: 
\begin{equation}
\text{MAE}=\dfrac{1}{N} \sum_{i=1}^N |y_i - x_i|
\end{equation}

\subsection{Mean Squared Error}
Mean squared error (MSE) is defined as the average of the squared difference between $\{x_i\}$ and $\{y_i\}$: 
\begin{equation}
\text{MSE}=\dfrac{1}{N} \sum_{i=1}^N (y_i - x_i)^2
\end{equation}



\clearpage
\section{Performance Assessment}

\subsection{MXMNet-MBE and PAMNet-MBE on Double-Density Datasets Using Original Training Strategy}

\begin{figure}[!h]
\centering
\includegraphics[width=0.5\textwidth]{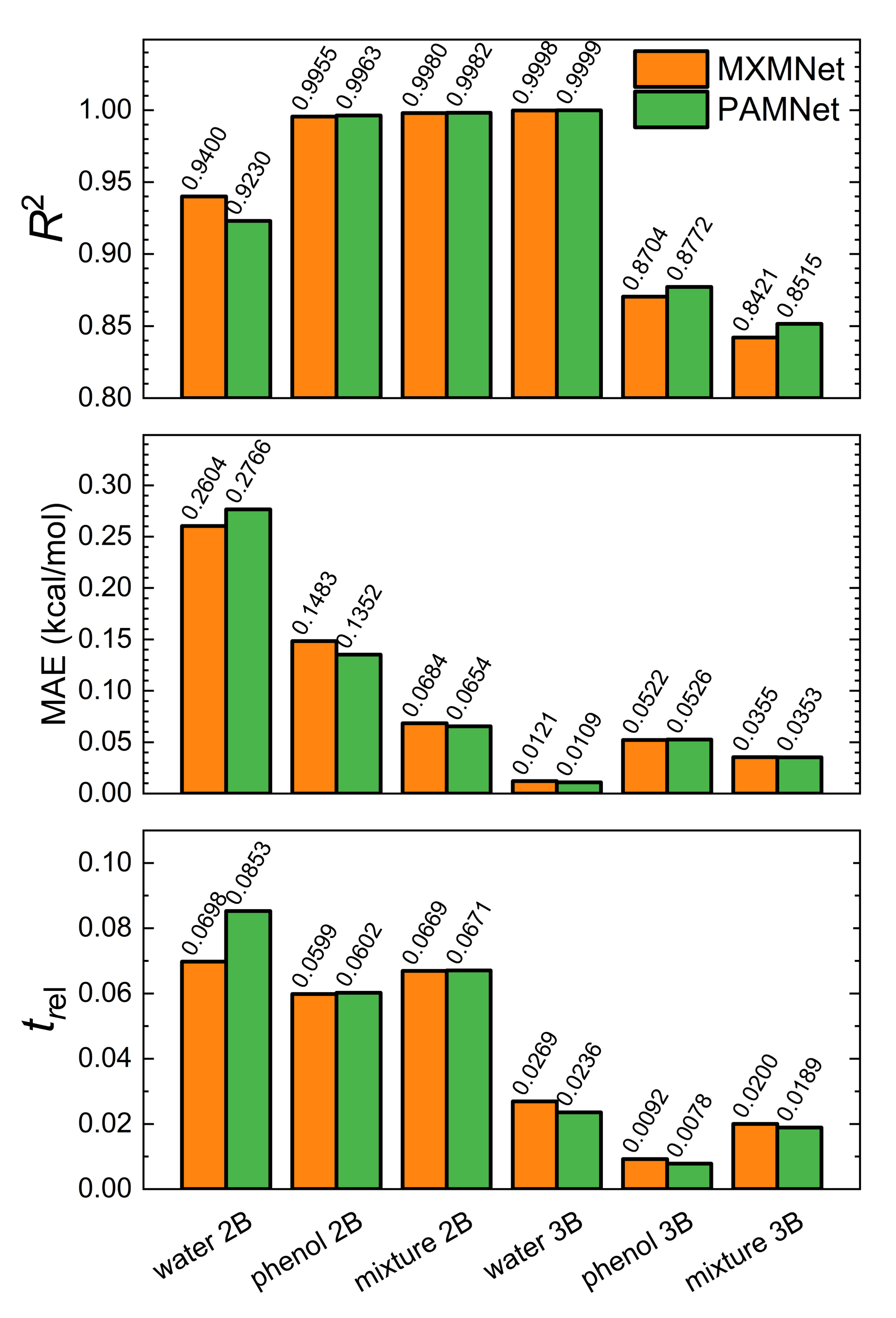}
\caption{Performance metrics of 2B and 3B energies on double-density clusters from MXMNet-MBE and PAMNet-MBE using the original training strategy in terms of $R^2$ (top), MAE (kcal/mol, middle), and $t_\text{rel}$ 
(bottom).} 
\label{fig:comparative_performance}
\end{figure}

The accuracy of MXMNet-MBE and PAMNet-MBE on double-density benchmark clusters using the original training strategy is summarized in Figures \ref{fig:MXMNet_PAMNet_results} and \ref{fig:comparative_performance} and Table \ref{tab:comparative_performance} and compared with non-FB-GNN-MBE methods in Table \ref{tab:comp_gnns}.

\begin{figure}[!h]
\centering
\includegraphics[width=0.5\textwidth]{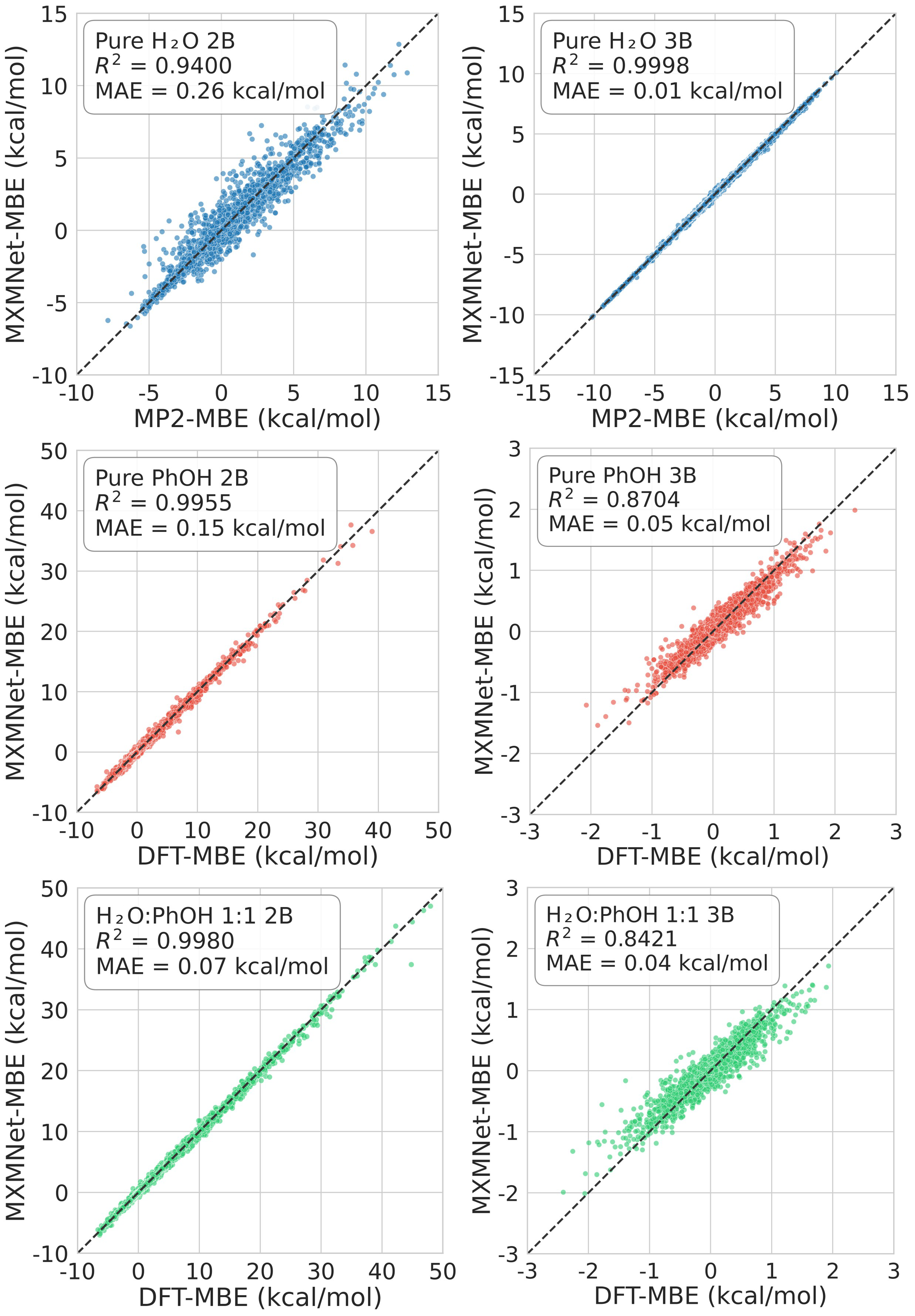}\includegraphics[width=0.5\textwidth]{Figures/Figure5b_PAMNet_results.png}
\caption{2B and 3B energies on double-density clusters from MXMNet-MBE (left) and PAMNet-MBE (right) using the original training strategy, compared with MP2-MBE or DFT-MBE.} 
\label{fig:MXMNet_PAMNet_results}
\end{figure}

\begin{table}[!h]
    \centering
    \caption{Performance metrics of 2B and 3B energies on double-density clusters from MXMNet-MBE and PAMNet-MBE using the original training strategy.} 
    \begin{tabular}{c|c|c|c|c|c|c|c}
    \hline\hline
        FB-GNN & dataset & $R^2$ & \multicolumn{1}{c|}{$\langle E_{\text{MP2/DFT}} \rangle$} & \multicolumn{1}{c|}{ME} & \multicolumn{1}{c|}{MAE} & \multicolumn{1}{c|}{$\langle t_{\text{MP2/DFT}} \rangle$} & \multicolumn{1}{c}{$\langle t_{\text{FB-GNN}} \rangle$}\\
        \cline{4-8}
        & & & \multicolumn{3}{c|}{(kcal/mol)} & \multicolumn{2}{c}{(s)} \\\hline\hline
        \multirow{6}{*}{MXMNet} & \ch{H2O} 2B & 0.9400 & $+$0.2550 & $+$0.0015 & 0.2604 & 1.29  & 0.09 \\ 
        & \ch{H2O} 3B & 0.9998 & $-$0.0008 & $+$0.0005 & 0.0121 & 2.97 & 0.08 \\ 
        & \ch{C6H5OH} 2B & 0.9955 & $+$1.1199 & $-$0.0002 & 0.1483 & 147.51 & 8.83  \\ 
        & \ch{C6H5OH} 3B & 0.8704 & $+$0.0010 & $+$0.0009 & 0.0522 & 428.17 & 3.94 \\ 
        & \ch{H2O:C6H5OH} 2B & 0.9980 & $+$0.8180 & $-$0.0044 & 0.0684 & 69.77 & 4.67 \\ 
        & \ch{H2O:C6H5OH} 3B & 0.8421 & $-$0.0088 & 0.0000 & 0.0355 & 198.77 & 3.98 \\ \hline
        \multirow{6}{*}{PAMNet} & \ch{H2O} 2B & 0.9230 & $+$0.2550 & $+$0.0021 & 0.2766 & 1.29 & 0.11 \\ 
        & \ch{H2O} 3B & 0.9999 & $-$0.0008 & $+$0.0015 & 0.0109 & 2.97 & 0.07 \\ 
        & \ch{C6H5OH} 2B & 0.9963 & $+$1.1199 & $-$0.0063 & 0.1352 & 147.51 & 8.88 \\ 
        & \ch{C6H5OH} 3B & 0.8772 & $+$0.0010 & $+$0.0002 & 0.0526 & 428.17 & 3.35 \\ 
        & \ch{H2O:C6H5OH} 2B & 0.9982 & $+$0.8180 & $-$0.0024 & 0.0654 & 69.77 & 4.68 \\ 
        & \ch{H2O:C6H5OH} 3B & 0.8515 & $-$0.0088 & $+$0.0001 & 0.0353 & 198.77 & 3.76 \\ \hline\hline
    \end{tabular}
    \label{tab:comparative_performance}
\end{table}

\begin{table}[!ht]
    \centering
    \caption{2B and 3B energies on double-density water clusters from FB-GNN-MBE and non-FB-GNN-MBE models using the original training strategy.}
    \begin{tabular}{c|c|c|c|c}
    \hline\hline
    dataset & \multicolumn{2}{c|}{\ch{H2O} 2B} & \multicolumn{2}{c}{\ch{H2O} 3B} \\\hline
    GNN & \multicolumn{1}{c|}{$R^2$} & \multicolumn{1}{c|}{MAE (kcal/mol)} & \multicolumn{1}{c|}{$R^2$} & \multicolumn{1}{c}{MAE (kcal/mol)} \\ \hline
    MXMNet & 0.9400 & 0.2604 & 0.9998 & 0.0121 \\ 
    PAMNet & 0.9230 & 0.2766 & 0.9999 & 0.0109 \\ 
    FragGen & 0.8053 & 0.5603 & 0.9851 & 0.2084\\ 
    MACE & 0.7750 & 0.4791 & 0.9616 & 0.3085 \\ 
    SchNet & 0.6491 & 0.8756 & 0.9783 & 0.0957\\ 
    DimeNet & 0.6638 & 0.8796 & 0.9958 & 0.0240 \\ 
    DimeNet++ & 0.6545 & 0.8698 & 0.9986 & 0.0214 \\ 
    ViSNet & 0.6532 & 0.8752 & 0.9991 & 0.0120 \\ \hline\hline
    \end{tabular}
    \label{tab:comp_gnns}
\end{table}

\clearpage

\subsection{PAMNet-MBE on Mixed-Density Clusters Using 
Multi-Stage Training Strategy}

\begin{figure}[!h]
\centering
\includegraphics[width=0.5\textwidth]{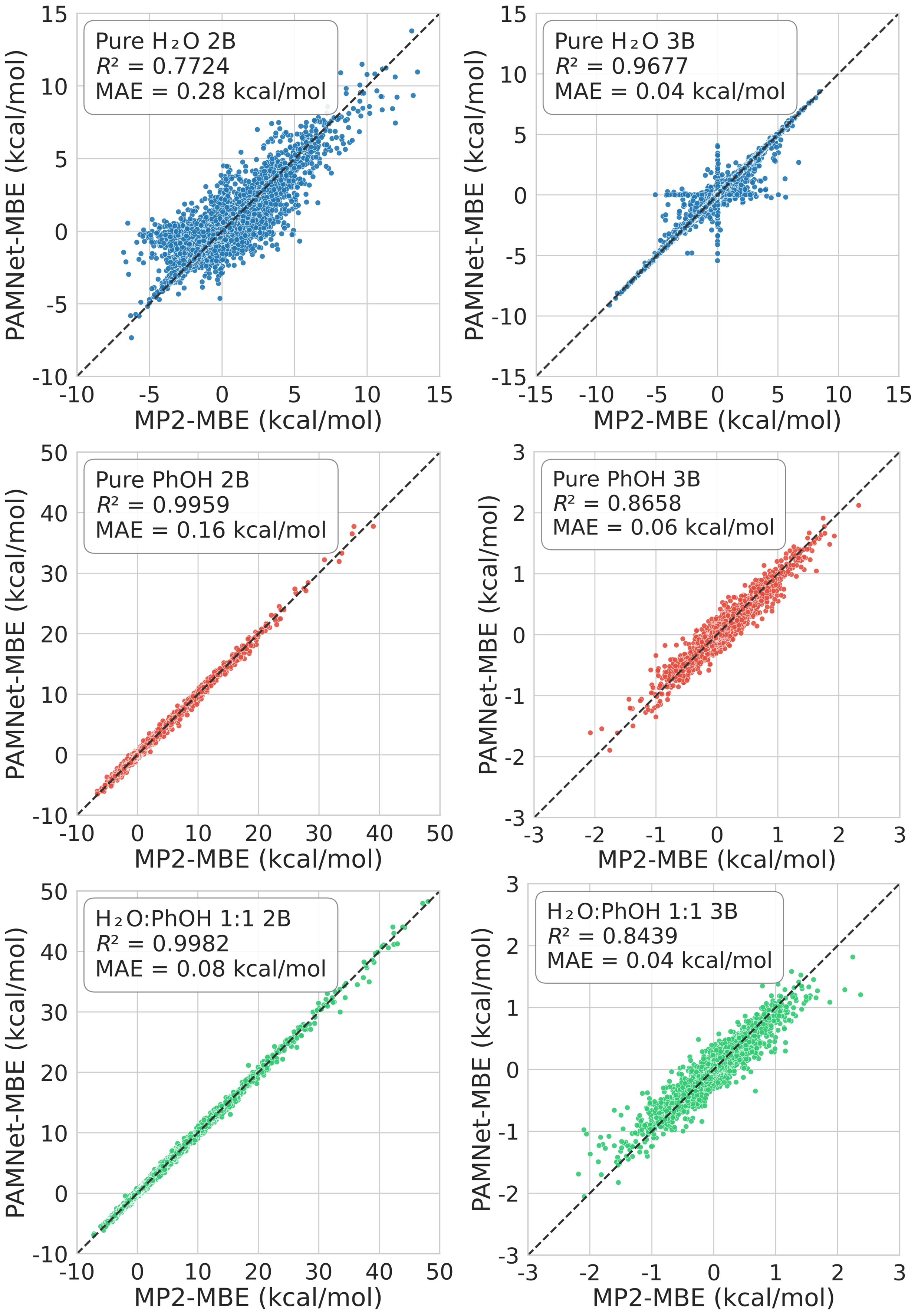}
\caption{2B and 3B energies on mixed-density water clusters and double-density phenol and water:phenol clusters from PAMNet-MBE using the multi-stage training strategy, compared with MP2-MBE or DFT-MBE.} 
\label{fig:multi_stage_results-1}
\end{figure}

\begin{figure}[!h]
\centering
\includegraphics[width=0.5\textwidth]{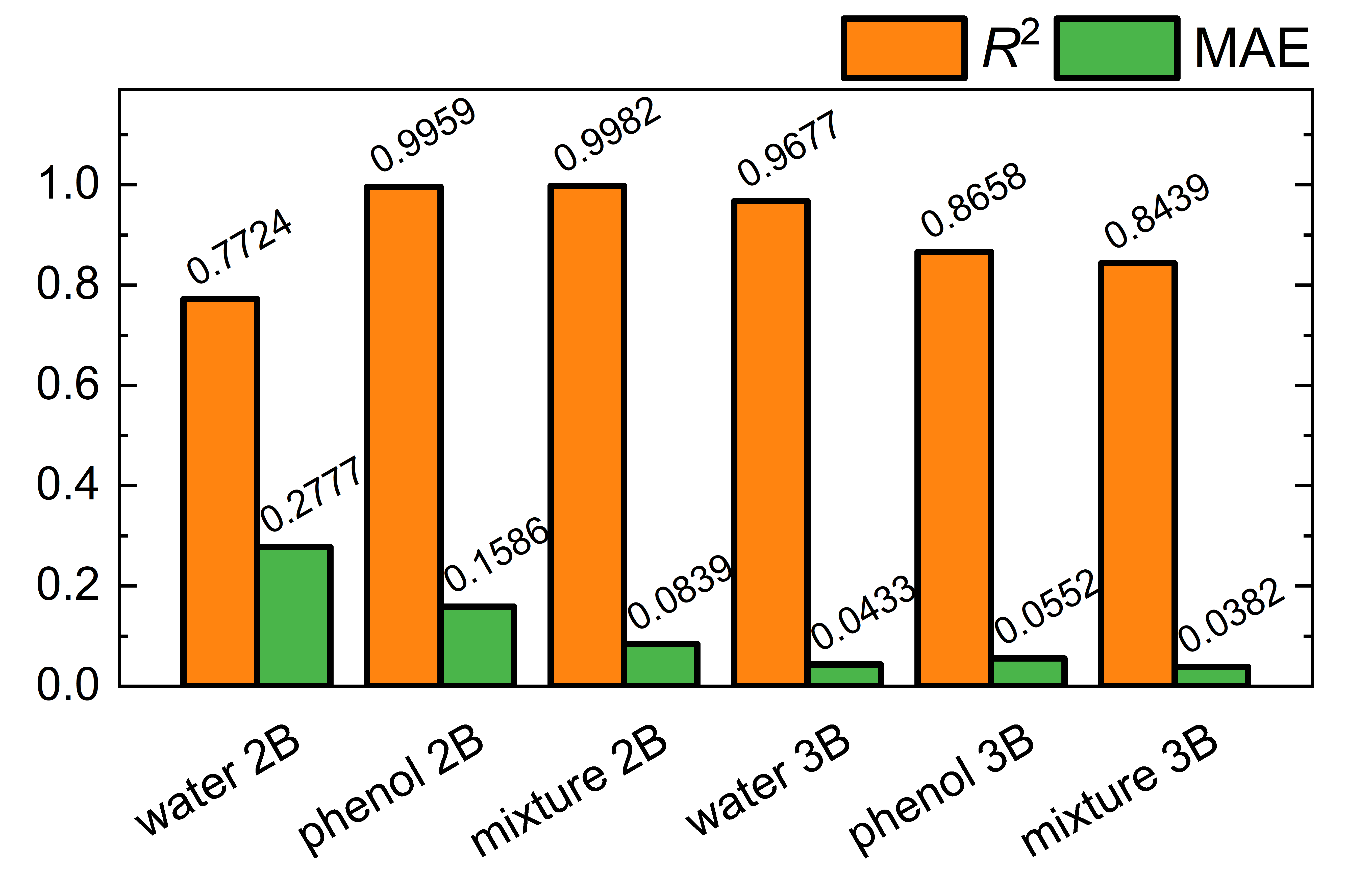}
\caption{Performance metrics of 2B and 3B energies on mixed-density water clusters and double-density phenol and water:phenol clusters from PAMNet-MBE using the multi-stage training strategy in terms of $R^2$ and MAE (kcal/mol).}
\label{fig:multi_stage_results-2}
\end{figure}

The accuracy of PAMNet-MBE on mixed-density water clusters and double-density phenol and water:phenol clusters using the multi-stage training strategy is summarized in Figures \ref{fig:multi_stage_results-1} and \ref{fig:multi_stage_results-2} and Table \ref{tab:multi_stage_performance}.


\begin{table}[!h]
    \centering
    \caption{Performance metrics of 2B and 3B energies on mixed-density water clusters and double-density phenol and water:phenol clusters from PAMNet-MBE using the multi-stage training strategy.} 
    \begin{tabular}{c|c|c|c}
    \hline\hline
    dataset & $R^2$ & ME (kcal/mol) & MAE (kcal/mol) \\
    \hline
    \ch{H2O} 2B & 0.7724 & $+0.0352$ & 0.2777 \\
    \ch{H2O} 3B & 0.9677 & $+0.0162$ & 0.0433 \\
    \ch{C6H5OH} 2B & 0.9959 & $-0.0156$ & 0.1586 \\
    \ch{C6H5OH} 3B & 0.8658 & $-0.0100$ & 0.0552 \\
    \ch{H2O}:\ch{C6H5OH} 2B & 0.9982 & $+0.0172$ & 0.0839 \\
    \ch{H2O}:\ch{C6H5OH} 3B & 0.8439 & $-0.0061$ & 0.0382 \\ 
    \hline\hline
    \end{tabular}
    \label{tab:multi_stage_performance}
\end{table}

\clearpage

\subsection{Fine-Tuned PAMNet-MBE on One-Dimensional Potential Energy Surfaces}

The accuracy of fine-tuned PAMNet-MBE on 1D PES (dimer dissociation curve) 
is summarized and compared with original MXMNet-MBE and PAMNet-MBE re-trained from scratch in Table \ref{tab:comp_pes}.

\begin{table}[!ht]
    \centering
    \caption{Performance metrics of 2B energies on 1D PES from original MXMNet-MBE, original PAMNet-MBE, and fine-tuned PAMNet-MBE.} 
    \begin{tabular}{c|c|c|c}
    \hline\hline
    dataset & \multicolumn{3}{c}{\ch{H2O} 2B} \\ \hline
   FB-GNN & $R^2$ & MAE (kcal/mol) & $t_\text{rel}$ \\ \hline\hline
    MXMNet (original)             & 0.3331 & 0.4918 & 0.1792 \\ 
    PAMNet (original)               & 0.7960 & 0.2324 & 1.4336 \\ 

    PAMNet (fine-tuned) 
    & 0.9685 & 0.0807 & 0.1887 \\ 
    \hline\hline
    \end{tabular}
    \label{tab:comp_pes}
\end{table}


\clearpage

\subsection{
Fine-Tuned Student Models on Unseen Water Clusters}
\label{sec:architecture}

\begin{figure}[!ht]
\centering
\includegraphics[width=0.5\textwidth]{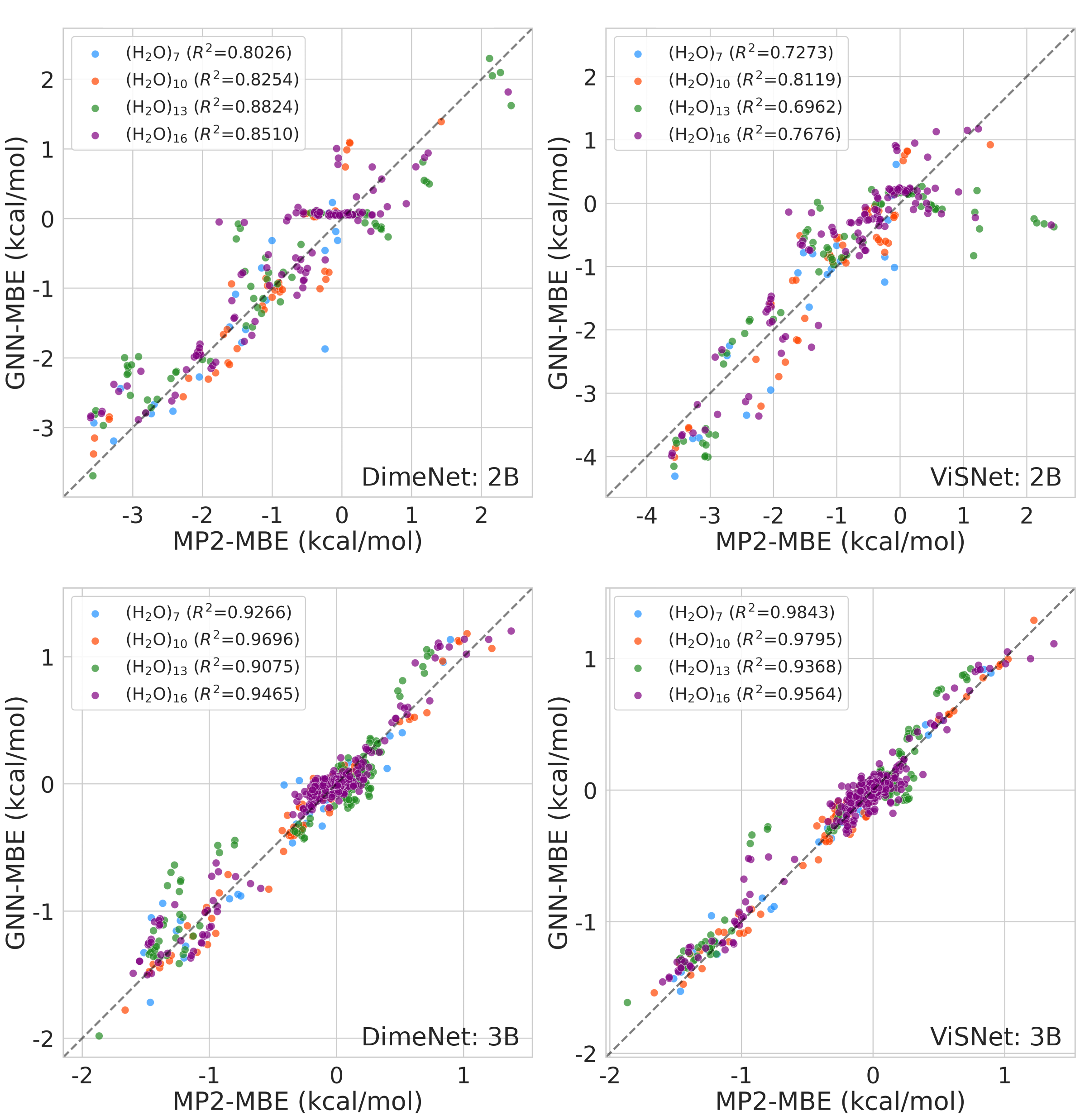}
\caption{2B and 3B energies on $\ch{(H2O)}_{7}$, $\ch{(H2O)}_{10}$, $\ch{(H2O)}_{13}$, and $\ch{(H2O)}_{16}$ clusters from DimeNet-MBE and ViSNet-MBE student models using the teacher--student knowledge distillation protocol, compared with MP2-MBE.} 
\label{fig:3}
\end{figure}

The accuracy of all fine-tuned student models on external fine-tuning sets of $\ch{(H2O)}_{21}$ clusters 
using the teacher--student distillation protocol 
is summarized in Table \ref{tab:finetune_best-question}.
Prediction performance on the unseen test set of smaller water clusters of $\ch{(H2O)}_{7}$, $\ch{(H2O)}_{10}$, $\ch{(H2O)}_{13}$, and $\ch{(H2O)}_{16}$ is summarized in Figure \ref{fig:3} and Tables \ref{tab:transfer_all_models-1}--
\ref{tab:comp_water_clusters}. 

This is the ultimate test of the practical utility of our FB-GNN-MBE models, in which we directly fine-tuned the student model ``taught'' by PAMNet to ensure high accuracy when adapting to new data. 
To capture the target energy landscape, we fine-tuned light-weight student models initialized from PAMNet and reused physically informed representations learned from the mixed-density water dataset and enabling reliable adaptation to the target system and allowing direct assessment whether the learned many-body physics transfers across cluster sizes. 
An adaptive learning rate schedule supports efficient refinement, while early stopping prevents overfitting to the limited configurations. 
As a result, the model learns the underlying shape of the PES and remains predictive on unseen structures.

\begin{table}[h]
\caption{Performance metrics of 2B and 3B energies on $\ch{(H2O)}_{21}$ clusters from the teacher model and four fine-tuned student models. Bold values denote the best student results.} 
\label{tab:finetune_best-question}
\begin{tabular}{c|c|c|c|c|c}
\hline\hline
dataset & model & MAE (kcal/mol) & $R^2$ & $\langle t_{\text{MP2}}\rangle$ (s) & $\langle t_{\text{ML}}\rangle$ (s) \\
\hline\hline
\multirow{4}{*}{2B} & PAMNet & 0.0637 & 0.9718 & 1.05 & 0.31 \\
& DimeNet & \textbf{0.2392} & \textbf{0.9030} & 1.83 & 1.65 \\
& DimeNet++ & 0.2601 & 0.8675 & 1.83 & 1.75 \\
& ViSNet & 0.3276 & 0.7301 & 1.83 & 1.33 \\
& SchNet & 0.4252 & 0.6520 & 1.83 & 0.25 \\
\hline\hline
\multirow{4}{*}{3B} & PAMNet & 0.0120 & 0.5903& 3.55 & 0.31 \\
&  DimeNet++ & \textbf{0.0263} & \textbf{0.9686} & 5.55 & 1.81 \\
& DimeNet & 0.0281 & 0.9554 & 5.55 & 1.26 \\
& ViSNet & 0.0298 & 0.9627 & 5.55 & 1.31 \\
& SchNet & 0.0658 & 0.6904 & 5.55 & 0.75 \\
\hline\hline
\end{tabular}
\end{table}

\begin{table}[h]
\caption{Performance metrics of 2B and 3B energies on $\ch{(H2O)}_{7}$, $\ch{(H2O)}_{10}$, $\ch{(H2O)}_{13}$, and $\ch{(H2O)}_{16}$ clusters four fine-tuned student models, two original teacher models, and one fine-tuned teacher model using the teacher--student knowledge distillation protocol.}
\label{tab:transfer_all_models-1}
\small \begin{tabular}{c|c|c|c|c|c|c|c|c}
\hline\hline
 dataset & cluster & \multicolumn{3}{c|}{DimeNet} & \multicolumn{3}{c|}{DimeNet++} & \\
 \hline
& & MAE & $R^2$ & $\langle t_{\text{GNN}}\rangle$ & MAE & $R^2$ & $ \langle t_{\text{GNN}}\rangle$ & $\langle t_{\text{MP2}}\rangle$ \\
 & & (kcal/mol) & & (s) & (kcal/mol) & & (s) & (s) \\
\hline\hline
\multirow{4}{*}{2B} &
$\ch{(H2O)}_{7}$  & 0.3443 & \textcolor{black}{0.8026} & 1.38 & 0.4350 & \textcolor{black}{0.7555} & 1.24 & 2.63\\
&$\ch{(H2O)}_{10}$ & 0.3606 & \textcolor{black}{0.8254} & 0.53 & 0.4033 & \textcolor{black}{0.7843} & 0.53 & 2.63\\
&$\ch{(H2O)}_{13}$ & 0.3851 & \textcolor{black}{0.8824} & 0.33 & 0.4595 & \textcolor{black}{0.8415} & 0.33 & 2.05 \\
&$\ch{(H2O)}_{16}$ & 0.3247 & \textcolor{black}{0.8510} & 0.20 & 0.3543 & \textcolor{black}{0.8243} & 0.19 & 1.92 \\
\hline\hline
\multirow{4}{*}{3B} & $\ch{(H2O)}_{7}$  & 0.1328 & \textcolor{black}{0.9266} & 0.71 & 0.1201 & \textcolor{black}{0.9396} & 0.77 & 8.87 \\
& $\ch{(H2O)}_{10}$ & 0.0638 & \textcolor{black}{0.9696} & 0.19 & 0.0791 & \textcolor{black}{0.9580} & 0.20 & 5.52 \\
& $\ch{(H2O)}_{13}$ & 0.0828& \textcolor{black}{0.9075} & 0.08 & 0.0853 & \textcolor{black}{0.8730} & 0.09 & 8.06 \\
& $\ch{(H2O)}_{16}$ & 0.0496 & \textcolor{black}{0.9465} & 0.05 & 0.0603 & \textcolor{black}{0.8963} & 0.05 & 6.58 \\
\hline\hline
\end{tabular}
\end{table}

\begin{table}[h]
\caption{Performance metrics of 2B and 3B energies on $\ch{(H2O)}_{7}$, $\ch{(H2O)}_{10}$, $\ch{(H2O)}_{13}$, and $\ch{(H2O)}_{16}$ clusters four fine-tuned student models, two original teacher models, and one fine-tuned teacher model using the teacher--student knowledge distillation protocol (continued).}
\label{tab:transfer_all_models-2}
\small \begin{tabular}{c|c|c|c|c|c|c|c|c}
\hline\hline
 dataset & cluster & \multicolumn{3}{c|}{ViSNet} & \multicolumn{3}{c|}{SchNet} & \\
 \hline
& & MAE & $R^2$ & $\langle t_{\text{GNN}}\rangle$ & MAE & $R^2$ & $ \langle t_{\text{GNN}}\rangle$ & $\langle t_{\text{MP2}}\rangle$ \\
 & & (kcal/mol) & & (s) & (kcal/mol) & & (s) & (s) \\
\hline\hline
\multirow{4}{*}{2B} &
$\ch{(H2O)}_{7}$  & 0.4928 & \textcolor{black}{0.7273} & 0.52 & 0.7381 & \textcolor{black}{0.0331} & 0.43 & 2.63\\
&$\ch{(H2O)}_{10}$ & 0.3934 & \textcolor{black}{0.8119} & 0.24 & 0.6490 & \textcolor{black}{0.4164} & 0.18 & 2.63\\
&$\ch{(H2O)}_{13}$ & 0.3582 & \textcolor{black}{0.6962} & 0.14 & 0.6657 & \textcolor{black}{0.6864} & 0.10 & 2.05 \\
&$\ch{(H2O)}_{16}$ & 0.4087 & \textcolor{black}{0.7676} & 0.09 & 0.6185 & \textcolor{black}{0.4501} & 0.07 & 1.92 \\
\hline\hline
\multirow{4}{*}{3B} & $\ch{(H2O)}_{7}$  & 0.0583 & \textcolor{black}{0.9843} & 0.34 & 0.3116 & \textcolor{black}{0.4709} & 0.23 & 8.87 \\
& $\ch{(H2O)}_{10}$ & 0.0561 & \textcolor{black}{0.9795} & 0.09 & 0.1872 & \textcolor{black}{0.7469} & 0.07 & 5.52 \\
& $\ch{(H2O)}_{13}$ & 0.0698 & \textcolor{black}{0.9368} & 0.05 & 0.1414 & \textcolor{black}{0.7021} & 0.03 & 8.06 \\
& $\ch{(H2O)}_{16}$ & 0.0459 & \textcolor{black}{0.9564} & 0.02 & 0.1047 & \textcolor{black}{0.7117} & 0.02 & 6.58 \\
\hline\hline
\end{tabular}
\end{table}


Architecture determines how well a student model generalizes to unseen water clusters. 
Since hydrogen bonds are highly directional, a distance-based model like SchNet generalizes less successfully. 
In contrast, a model that explicitly encodes geometric information performs better.  
DimeNet (using bond angles) and ViSNet (using vectorized directions) successfully capture the nature of intermolecular interactions in water. 
Their high performance on unseen water clusters (even with limited fine-tuning) confirms that the teacher--student knowledge distillation protocol successfully transfers knowledge from the general domain to the specific application domain.

Compared to DimeNet, ViSNet produces a broader distribution of small-magnitude 2B energies rather than concentrating them near zero. 
This behavior leads to a more uniformly populated parity plot around the diagonal line in the low-energy regime. 
However, this increased dispersion also results in larger absolute deviations from MP2-calculated energies, reflected in slightly higher MAE values reported in Tables \ref{tab:transfer_all_models-1} and \ref{tab:transfer_all_models-2}. 
For 3B energies, ViSNet achieves the best performance across all four unseen clusters, yielding the lowest MAE and the highest $R^2$ values up to \textcolor{black}{0.9843}. 
This ranking differs from the fine-tuning results of $\ch{(H2O)_{21}}$ for which DimeNet++ performs best, indicating that optimal predictions of 3B energies depend on the target structural regime rather than on in-distribution performance alone. 
DimeNet++ benefits from an angular basis that closely matches the local environments present in $\ch{(H2O)}_{21}$, whereas ViSNet generalizes more smoothly when the hydrogen bond network and coordination patterns change. 
As a result, ViSNet provides more consistent predictions of 3B energies across cluster sizes, highlighting its advantage for transferable many-body modeling.

\begin{table}[ht]
\centering
\caption{Performance metrics of 2B and 3B energies on $\ch{(H2O)}_{7}$, $\ch{(H2O)}_{10}$, $\ch{(H2O)}_{13}$, and $\ch{(H2O)}_{16}$ clusters from two original teacher models, one fine-tuned teacher model, and four fine-tuned students models using the teacher--student knowledge distillation protocol (continued-2).} 
\label{tab:comp_water_clusters}
\resizebox{\textwidth}{!}{%
\begin{tabular}{c|c|c|c|c|c|c|c|c|c|c|c}
\hline\hline
dataset & cluster &
\multicolumn{3}{c|}{original MXMNet} &
\multicolumn{3}{c|}{original PAMNet} &
\multicolumn{3}{c|}{fine-tuned PAMNet} & \\ 
\hline
 &  & MAE & $R^{2}$ & $\braket{t_\text{GNN}}$ & MAE & $R^{2}$ & $\braket{t_\text{GNN}}$ & MAE & $R^{2}$ & $\braket{t_\text{GNN}}$ & $\braket{t_\text{MP2}}$ \\
 & & (kcal/mol) & & (s) & (kcal/mol) & & (s) & (kcal/mol) & & (s) & (s) \\
\hline\hline
\multirow{4}{*}{2B} 
& $(\ch{H2O})_{7}$  & 1.7231 & $-$3.4135 & 48.52 & 2.2031 & $-$5.1589 & 8.05 & 1.0646 & $-$0.5175 & 0.90 & 2.63 \\ 
& $(\ch{H2O})_{10}$ & 0.4180 & 0.8488 & 26.62 & 0.9310 & 0.2792 & 25.39 & 0.8501 & $-$0.4112 & 0.58 & 2.63 \\ 
& $(\ch{H2O})_{13}$ & 0.5914 & 0.8098 & 15.29 & 1.2119 & 0.0486 & 15.63 & 0.4478 & 0.8519 & 0.33 & 2.05 \\ 
& $(\ch{H2O})_{16}$ & 0.4666 & 0.6807 & 5.98 & 0.3599 & 0.8010 & 11.32 & 0.3735 & 0.7509 & 0.33 & 1.92 \\
\hline\hline
\multirow{4}{*}{3B} 
& $(\ch{H2O})_{7}$  & 1.5947 & $-$4.2383 & 40.14 & 0.5010 & 0.2926 & 36.46 & 1.3592 & $-$4.3703 & 0.86 & 8.87 \\ 
& $(\ch{H2O})_{10}$ & 0.1895 & 0.6697 & 8.73 & 0.2586 & 0.3116 & 6.85 & 0.8125 & $-$4.0628 & 0.31 & 5.52 \\
& $(\ch{H2O})_{13}$ & 0.3259 & 0.3112 & 1.88 & 0.1304 & 0.8631 & 5.23 & 0.2504 & 0.4494 & 0.36 & 8.06 \\
& $(\ch{H2O})_{16}$ & 0.1628 & 0.6973 & 0.70 & 0.2108 & $-$0.0898 & 0.73 & 0.1405 & 0.6364 & 0.30 & 6.58 \\
\hline\hline
\end{tabular}}
\end{table}

\clearpage

\section{Uniform Manifold Approximation and Projection}

\begin{figure}[h]
\centering
\includegraphics[width=\textwidth]{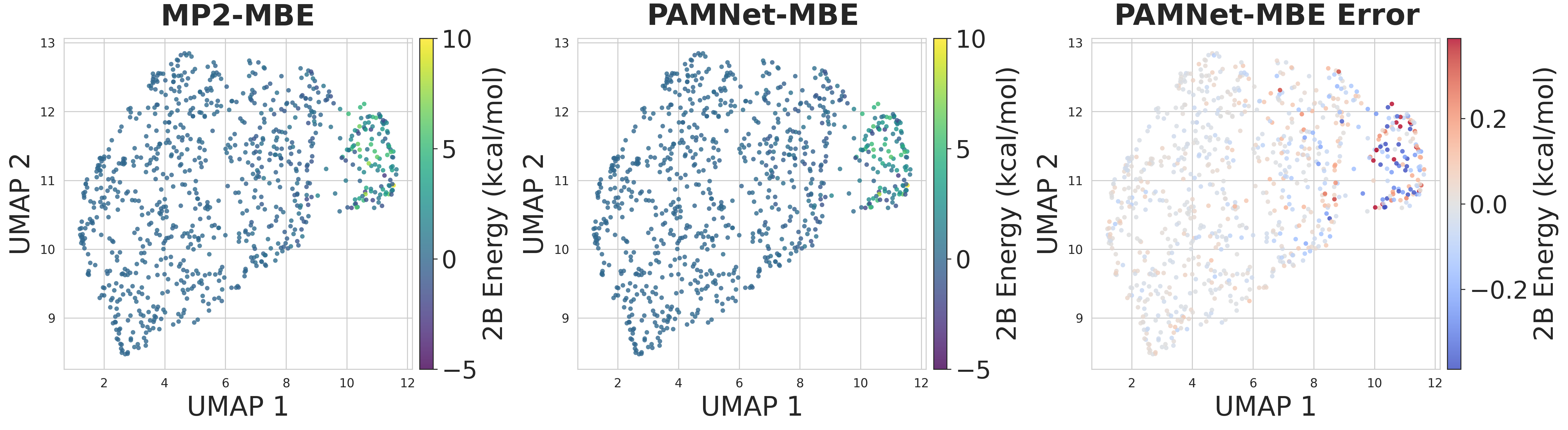}
\caption{UMAP visualization of the learned latent space for 2B energies, with MP2-calculated values (left), PAMNet-predicted values (center), and prediction errors (right).}
\label{fig:2b_umap}
\end{figure}

To further validate the learning capacity of FB-GNN-MBE beyond simple scalar metrics, we employ uniform manifold approximation and projection (UMAP) to visualize the system features 
into 2D {for double-density water clusters}. \cite{mcinnes2018umap} 
The results show characters for 2B interactions and 
help interpret the error statistics (Figure \ref{fig:2b_umap}). 
The latent space forms a single, extended manifold.
This is consistent with the wide range of pairwise 2B energies, which covers short-range repulsions, medium-range attractions, and long-range dissociations. 
Prediction errors are spread across the manifold and alternate in sign, without forming large contiguous regions of 
overestimation or underestimation. 

\clearpage
\section{Hyperparameters and Sensitivity Analysis for Water Clusters}

\subsection{Key and Optimized Hyperparameters}
The training process begins with pre-training of PAMNet-MBE using the multi-stage training strategy. 
The learning rate is reduced at each stage (lr\_s1 $>$ lr\_s2 $>$ lr\_s3), which allows the network to learn strong interaction patterns early and gradually refine its predictions in low-energy regions. 
The pre-trained result is then used in two applications. 
First, it serves as a teacher model for the knowledge distillation protocol, in which physical representations learned by PAMNet are transferred to simpler GNN architectures (DimeNet, DimeNet++, SchNet, and ViSNet). 
A weighted feature loss controls the emphasis placed on matching internal representations versus final energy predictions, and performance is evaluated on unseen molecular clusters. 
Second, PAMNet-MBE is directly fine-tuned on small target datasets to test whether the pre-trained representations can reproduce the energy landscape with limited data. 
Fine-tuning is performed with a low learning rate (ft\_lr) to adapt the model without disrupting previously learned physical trends. 
The key hyperparameters for the multi-stage training strategy and the knowledge distillation protocol, along with their optimized values, are summarized in Tables \ref{tab:hparams_staged}--\ref{tab:finetune_best}.




\begin{table}[h]
\centering
\caption{Key hyperparameters for multi-stage training.}
\label{tab:hparams_staged}
\begin{tabular}{c|c|c|c|c} 
\hline\hline
hyperparameter & default & alternative & part & note \\
\hline\hline
n\_dim & 128 & 64, 256 & [Architecture] & hidden dimension of PAMNet \\
\hline
n\_layer & 5 & 3, 4, 6 & [Architecture] & number of message passing layers \\
\hline
cutoff\_l & 5.0 & 1.7 & [Architecture] & local interaction cutoff (\AA) \\
\hline
cutoff\_g & 10.0 & 5.0 & [Architecture] & global interaction cutoff (\AA) \\
\hline
batch\_size & 64 & 32, 128 & [Training] & graphs processed per training step\\
\hline
patience & 20 & 15, 30 & [Training] & early stopping patience (epochs) \\
\hline
$\tau_{\text{high}}$ & $P_{75}(|E|)$ & - & [Stage 1] & energy threshold for high-energy data \\
\hline
lr\_s1 & $5 \times 10^{-5}$ & $1\times 10^{-4}$ & [Stage 1] & learning rate for Stage 1 \\
\hline
epochs\_s1 & 300 & 200 & [Stage 1] & maximum epochs for Stage 1 \\
\hline
$\tau_{\text{med}}$ & $P_{50}(|E|)$ & - & [Stage 2] & energy threshold for medium-energy data \\
\hline
lr\_s2 & $3 \times 10^{-5}$ & $8 \times 10^{-5}$ & [Stage 2] & learning rate for Stage 2 \\
\hline
$\text{epochs\_s2}$ & 300 & 200 & [Stage 2] & maximum epochs for Stage 2 \\
\hline
lr\_s3 & $1 \times 10^{-5}$ & $5 \times 10^{-5}$ & [Stage 3] & learning rate for Stage 3 (full data) \\
\hline
$\text{epochs\_s3}$ & 300 & 200 & [Stage 3] & maximum epochs for Stage 3 \\
\hline\hline
\end{tabular}
\end{table}

\begin{table}[h]
\centering
\caption{Optimized hyperparameters 
for multi-stage training.}
\begin{tabular}{c|c|c}
\hline\hline
hyperparameter 
& 2B 
& 3B \\ 
\hline\hline
lr\_s1
& $1 \times 10^{-4}$ 
& $5 \times 10^{-5}$ \\
lr\_s2
& $8 \times 10^{-5}$ 
& $3 \times 10^{-5}$ \\
lr\_s3 
& $5 \times 10^{-5}$ 
& $1 \times 10^{-5}$ \\
\hline\hline
\end{tabular}
\label{tab:multi-stage-hyperparameters}
\end{table}




\begin{table}[h]
\centering
\caption{Key hyperparameters for teacher--student knowledge distillation.}
\label{tab:hparams_distill}
\begin{tabular}{c|c|c|c|c} 
\hline\hline
hyperparameter & default & alternative & part & note \\
\hline\hline
n\_dim & 64 & 128, 256 & [Architecture] & student hidden dimension size \\
\hline
n\_layer & 3 & 2, 4 & [Architecture] & student message passing layers\\
\hline
batch\_size & 16 & 32, 64 & [Training] & graphs processed per training step \\
\hline
distill\_lr & $1 \times 10^{-4}$ & $1 \times 10^{-5}$ & [Distillation] & learning rate for distillation phase\\
\hline
distill\_epochs & 150 & 100, 200 & [Distillation] & epochs for distillation phase \\
\hline
$\lambda$ & 0.01 & 0.00, 0.10 & [Distillation] & feature loss weight, 0 = no feature matching \\
\hline
distill\_layer & $-1$ & $0$, $1$, \dots & [Distillation] & layer for feature matching, $-1$ = final layer \\
\hline
ft\_lr & $1 \times 10^{-4}$ & $1 \times 10^{-5}$ & [Fine-tuning] & learning rate for fine-tuning phase\\
\hline
ft\_epochs & 200 & 100, 500 & [Fine-tuning] & epochs for fine-tuning on true labels\\
\hline\hline
\end{tabular}
\end{table}

\begin{table}[h]
\caption{Optimized hyperparameters for teacher--student knowledge distillation.}
\label{tab:finetune_best}
\begin{tabular}{c|c|c|c|c|c}
\hline\hline
term & student & n\_dim & n\_layer & $\lambda$ & ft\_lr \\
\hline\hline
\multirow{4}{*}{2B} & DimeNet & 64 & 3 & 0.01 & $1\times10^{-4}$ \\
& DimeNet++ & 128 & 3 & 0.01 & $1\times10^{-4}$ \\
& ViSNet & 128 & 3 & 0.01 & $1\times10^{-4}$ \\
& SchNet & 128 & 3 & 0.01 & $1\times10^{-4}$ \\
\hline
\multirow{4}{*}{3B} & DimeNet++ & 128 & 3 & 0.01 & $1\times10^{-4}$ \\
& DimeNet & 128 & 3 & 0.05 & $1\times10^{-4}$ \\
& ViSNet & 128 & 3 & 0.00 & $1\times10^{-4}$ \\
& SchNet & 128 & 3 & 0.00 & $1\times10^{-4}$ \\
\hline\hline
\end{tabular}
\end{table}

\subsection{Hyperparameter Sensitivity Analysis of Teacher--Student Knowledge Distillation Protocols}
\label{sec:analysis}

Here we assess how the hyperparameters affect the transferability of the teacher--student knowledge distillation protocol to 
unseen clusters 
(Tables \ref{tab:sensitivity_2b} and \ref{tab:sensitivity_3b}). 
We aim to find an optimal balance 
between model fidelity and computational efficiency.

\subsubsection{Hidden Dimension (n\_dim)}
We observe different effects of the hidden dimension in 2B and 3B energies. 
For 2B energies (Table \ref{tab:sensitivity_2b}), smaller models often perform better: 
DimeNet reaches its best MAE of 0.2392 kcal/mol when n\_dim $=64$, 
while a larger MAE of 0.2553 kcal/mol is observed when n\_dim $=128$. 
For 3B energies, in contrast (Table \ref{tab:sensitivity_3b}), DimeNet benefits from a higher model capacity and achieves its best MAE of 0.0281 kcal/mol when n\_dim $=128$, and 
ViSNet is more parameter-efficient because n\_dim $=64$ slightly outperforms n\_dim $=128$ (MAE of 0.0302 vs 0.0316 kcal/mol) while using fewer parameters.
Overall, 
n\_dim values between 64 and 128 yield the best performance.

\subsubsection{Number of Layers (n\_layer)}
A shallower network with a smaller n\_layer yields more robust transferability for the selected cluster. For 2B energies (Table \ref{tab:sensitivity_2b}), increasing DimeNet layers from 3 to 4 degrades the validation MAE from 0.2392 to 0.2913 kcal/mol. 
Reducing to 2 layers also hurts performance and increases MAE to 0.3241.
This result indicates that n\_layer $=3$ is the sweet spot for capturing sufficient intermolecular interactions without the optimization difficulty. 
This trend holds for 3B energies (Table \ref{tab:sensitivity_3b}), where increasing DimeNet layers from 3 to 4 yields no gain. 
In both cases, n\_layer $=3$ is sufficient for modeling these molecular clusters.

\subsubsection{Feature Loss Weight ($\lambda$)}
Feature distillation has a clear but architecture-dependent effect (Table \ref{tab:feat_loss_ablation}).
For 2B energies, adding a small feature loss $\lambda=0.01$ consistently improves performance for DimeNet++, ViSNet, SchNet by reducing their MAEs.  
DimeNet remains the exception by performing best without the feature loss.
For 3B energies, 
the impact of feature loss is more nuanced and capacity-dependent. 
Angle-aware models benefit more strongly from feature guidance. 
In DimeNet, increasing the feature loss to $\lambda=0.05$ gives the best result (Table \ref{tab:sensitivity_3b}). DimeNet++ shows a modest gain 
with mild guidance ($\lambda=0.01$). 
Interestingly, ViSNet exhibits a distinct ``capacity crossover'' effect. For the compact model (n\_dim $=64$), feature loss provides a massive performance boost, 
likely compensating for its limited expressivity. 
However, for the larger model (n\_dim $=128$), the unconstrained baseline performs better, implying that feature matching may over-regularize the model when it already has sufficient capacity to fit the surface. 
SchNet performs worse when feature loss is applied 
(MAE increasing from 0.0658 at $\lambda=0.00$ to 0.0818 at $\lambda=0.01$).

\subsubsection{Fine-Tuning Learning Rates (ft\_lr)}
ft\_lr 
is the dominant stability factor. 
Across all models, ft\_lr $=1\times 10^{-4}$ consistently yields the best results. 
Lowering ft\_lr to $1\times10^{-5}$ results in severe underfitting or convergence problem, as seen in the 2B energies of SchNet (MAE $=2.6110$ kcal/mol and $R^2 < 0$) and DimeNet (MAE $=0.3822$ kcal/mol). 
This strong performance at ft\_lr $=1\times 10^{-4}$ suggests that the distilled weights serve as a robust initialization, allowing effective fine-tuning toward the student’s target potential.

\subsubsection{Cutoff Distances}
cutoff\_l and cutoff\_g (in \AA) specify the cutoff distances beyond which the PAMNet architecture to ignore local and global interactions. 
Small cutoff distances (cutoff\_l $=1.7$ \AA\ and cutoff\_g $=5.0$ \AA) %
yield accurate 2B and 3B energies but lack 
physical validity. 
Especially, cutoff\_l $=1.7$ \AA\ is insufficient to capture the complete solvation shell required to describe the formation and destruction of hydrogen bonds.\cite{laage2017water}
By increasing the cutoff distances to cutoff\_l $=5.0$ \AA\ and cutoff\_g $=10.0$ \AA, the model can now account for the complex physics of hydrogen bond networks and long-range intermolecular interactions.

\subsubsection{Architectural Effects and Efficiency}
Among the models tested, DimeNet++ achieves the lowest MAE of 0.0263 kcal/mol in 3B energies. 
For 2B energies, 
where efficiency is more important, DimeNet with n\_dim $=64$ provides the best balance, offering the best accuracy--efficiency trade-off (Table \ref{tab:finetune_best}).
SchNet is the computationally least expensive but consistently shows higher errors, indicating that angular information is important for accurately modeling the PES of water clusters.

\begin{table}[h]
\centering
\caption{Sensitivity analysis on 2B energies from the $\ch{(H2O)}_{21}$ cluster. Bold values denote the best combination of hyperparameters.}
\label{tab:sensitivity_2b}
\begin{tabular}{c|c|c|c|c|c|c|c|c|c}
\hline\hline
{model} & n\_dim & n\_layer & \# params & $\lambda$ & ft\_lr & train MAE & val MAE & val $R^2$ & time \\
& & & (kilo) & & & (kcal/mol) & (kcal/mol) & & (min) \\
\hline\hline
\multirow{8}{*}{DimeNet}
 & 32  & 3 & 77   & 0.01 & $1\times10^{-4}$ & 0.1497 & 0.3558 & 0.6731 & 5.6 \\
 & 64  & 2 & 208  & 0.01 & $1\times10^{-4}$ & 0.1913 & 0.3241 & 0.8191 & 5.0 \\
 & \textbf{64}  & \textbf{3} & \textbf{292}  & \textbf{0.01} & \textbf{$1\times10^{-4}$} & \textbf{0.1353} & \textbf{0.2392} & \textbf{0.9030} & \textbf{5.2} \\
 & 64  & 4 & 376  & 0.01 & $1\times10^{-4}$ & 0.1546 & 0.2913 & 0.8236 & 5.8 \\
 & 128 & 3 & 1140 & 0.00 & $1\times10^{-4}$ & 0.1826 & 0.2553 & 0.8733 & 5.4 \\
 & 128 & 3 & 1140 & 0.01 & $1\times10^{-4}$ & 0.1380 & 0.2754 & 0.8948 & 5.9 \\
 & 128 & 3 & 1140 & 0.10 & $1\times10^{-4}$ & 0.1207 & 0.2571 & 0.8791 & 5.3 \\
 & 128 & 3 & 1140 & 0.05 & $1\times10^{-4}$ & 0.1431 & 0.2616 & 0.8855 & 5.7 \\
\hline
\multirow{5}{*}{DimeNet++}
 & 64  & 3 & 1026 & 0.01 & $1\times10^{-4}$ & 0.1552 & 0.2671 & 0.8754 & 5.2 \\
 & \textbf{128} & \textbf{3} & \textbf{1521} & \textbf{0.01} & \textbf{$1\times10^{-4}$} & \textbf{0.1503} & \textbf{0.2601} & \textbf{0.8675} & \textbf{5.5} \\
 & 128 & 3 & 1521 & 0.00 & $1\times10^{-4}$ & 0.1394 & 0.2684 & 0.8643 & 5.1 \\
 & 128 & 3 & 1521 & 0.01 & {$1\times10^{-5}$} & 0.1732 & 0.3189 & 0.7788 & 5.3 \\
 & 32  & 3 & 877    & 0.10 & $1\times10^{-4}$ & 0.1763 & 0.3288 & 0.7746 & 2.9 \\
\hline
\multirow{4}{*}{ViSNet}
 & 64  & 3 & 234  & 0.01 & $1\times10^{-4}$ & 0.1986 & 0.3284 & 0.7432 & 4.1 \\
 & \textbf{128} & \textbf{3} & \textbf{896}  & \textbf{0.01} & \textbf{$1\times10^{-4}$} & \textbf{0.1696} &  \textbf{0.3276} & \textbf{0.7301} & \textbf{4.2} \\
 & 128 & 3 & 896  & 0.00 & $1\times10^{-4}$ & 0.1907 & 0.3436 & 0.7588 & 4.0 \\
 & 128 & 3 & 896  & 0.01 & $1\times10^{-5}$ & 0.2940 & 0.4706 & 0.4707 & 4.2 \\
\hline
\multirow{4}{*}{SchNet}
 & \textbf{64} & \textbf{3} & \textbf{256}  & \textbf{0.01} & \textbf{$1\times10^{-4}$} & \textbf{0.3861} & \textbf{0.4252} & \textbf{0.6520} & \textbf{0.8} \\
 & 128 & 3 & 256  & 0.00 & $1\times10^{-4}$ & 0.1825 & 0.4541 & 0.6845 & 2.5 \\
 & 128 & 3 & 256  & 0.01 & $1\times10^{-4}$ & 0.2185 & 0.4419 & 0.6802 & 2.5 \\
 & 128 & 3 & 256  & 0.01 & $1\times10^{-5}$ & 2.3319 & 2.6110 & $-7.2880$ & 2.4 \\
\hline\hline
\end{tabular}
\end{table}

\begin{table}[h]
\centering
\caption{Sensitivity analysis on 3B energies from the $\ch{(H2O)}_{21}$ cluster. Bold values denote the best combination of hyperparameters.}
\label{tab:sensitivity_3b}
\begin{tabular}{c|c|c|c|c|c|c|c|c|c}
\hline\hline
{model} & n\_dim & n\_layer & \# params & $\lambda$ & ft\_lr & train MAE & val MAE & val $R^2$ & time \\
& & & (kilo) & & & (kcal/mol) & (kcal/mol) & & (min) \\
\hline\hline
\multirow{6}{*}{{DimeNet}}
 & 64  & 3 & 292  & 0.01 & $1\times10^{-4}$ & 0.0294 & 0.0300 & 0.9564 & 18.8 \\
 & 64  & 4 & 376  & 0.01 & $1\times10^{-4}$ & 0.0332 & 0.0312 & 0.9540 & 28.3 \\
 & 128 & 3 & 1140 & 0.00 & $1\times10^{-4}$ & 0.0322 & 0.0301 & 0.9541 & 32.4 \\
 & 128 & 3 & 1140 & 0.01 & $1\times10^{-4}$ & 0.0250 & 0.0283 & 0.9673 & 21.9 \\
 & \textbf{128} & \textbf{3} & \textbf{1140} & \textbf{0.05} & \textbf{$1\times10^{-4}$} & \textbf{0.0296} & \textbf{0.0281} & \textbf{0.9554} & \textbf{25.2} \\
 & 128 & 3 & 1140 & 0.01 & $1\times10^{-5}$ & 0.0570 & 0.0640 & 0.9026 & 30.6 \\
\midrule
\multirow{3}{*}{{DimeNet++}}
 & 64  & 3 & 1026 & 0.01 & $1\times10^{-4}$ & 0.0355 & 0.0345 & 0.9445 & 25.7 \\
 & 128 & 3 & 1521 & 0.00 & $1\times10^{-4}$ & 0.0264 & 0.0267 & 0.9691 & 28.1 \\
 & \textbf{128} & \textbf{3} & \textbf{1521} & \textbf{0.01} & \textbf{$1\times10^{-4}$} & \textbf{0.0281} & \textbf{0.0263} & \textbf{0.9686} & \textbf{36.2} \\
\midrule
\multirow{2}{*}{{ViSNet}}
 & 64  & 3 & 234  & 0.01 & $1\times10^{-4}$ & 0.0309 & 0.0302 & 0.9578 & 23.6 \\
 & \textbf{128} & \textbf{3} & \textbf{896} & \textbf{0.00} & \textbf{$1\times10^{-4}$} & \textbf{0.0325} & \textbf{0.0298} & \textbf{0.9627} & \textbf{26.1} \\
 & 128 & 3 & 896  & 0.01 & $1\times10^{-4}$ & 0.0330 & 0.0316 & 0.9448 & 20.9 \\
\midrule
\multirow{3}{*}{\textbf{SchNet}}
 & 64  & 3 & 163  & 0.01 & $1\times10^{-4}$ & 0.0592 & 0.0681 & 0.8470 & 14.3 \\
 & \textbf{128} & \textbf{3} & \textbf{256}  & \textbf{0.00} & \textbf{$1\times10^{-4}$} & \textbf{0.0679} & \textbf{0.0658} & \textbf{0.6904} & \textbf{15.0} \\
 & 128 & 3 & 256  & 0.01 & $1\times10^{-4}$ & 0.0719 & 0.0818 & 0.4997 & 15.1 \\
\hline\hline
\end{tabular}
\end{table}

\begin{table}[h]
\centering
\caption{Ablation study on impact of feature distillation loss ($\lambda$) on prediction performance. $\Delta$ shows the relative change of MAE from $\lambda=0.00$ to $0.01$.  Bold values denote the best combination of hyperparameters.}
\label{tab:feat_loss_ablation}
\begin{tabular}{c|c|c|c|c|c|c}
\hline\hline
dataset & model & n\_dim & n\_layer & \multicolumn{2}{c|}{MAE (kcal/mol)} & $\Delta$ (\%) \\
& & & & $\lambda=0.00$ &  $\lambda=0.01$ &  \\
\hline\hline
\multirow{4}{*}{2B} 
 & DimeNet    & 128 & 3 & \textbf{0.2553} & 0.2754 & $+7.8$ \\
 & DimeNet++  & 128 & 3 & 0.2684 & \textbf{0.2601} & $-3.1$ \\
 & ViSNet     & 128 & 3 & 0.3436 & \textbf{0.3276} & $-4.7$ \\
 & SchNet     & 128 & 3 & 0.4541 & \textbf{0.4419}  & $-2.7$ \\
\midrule
\multirow{5}{*}{3B} 
 & DimeNet    & 128 & 3 & 0.0301 & \textbf{0.0283} & $-6.0$ \\
 & DimeNet++  & 128 & 3 & 0.0267 & \textbf{0.0263} & $-1.5$ \\
 & SchNet     & 128 & 3 & \textbf{0.0658} & 0.0818 &  $+24.3$ \\
 & ViSNet     & 128 & 3 & \textbf{0.0298} & 0.0316 & $+6.0$ \\
\hline\hline
\end{tabular}
\end{table}

\clearpage






\bibliography{cite.bib}

\end{document}